\documentclass[%
 reprint,
amsmath,amssymb,
aps,
pra,
]{revtex4-1}
\usepackage{graphicx}% Include figure files
\usepackage{dcolumn}% Align table columns on decimal point
\usepackage{bm}% bold math
\usepackage{multirow}
\usepackage[dvipdfm,colorlinks,linkcolor=blue, urlcolor=blue, anchorcolor=blue, citecolor=blue]{hyperref}
\begin{document}

%\preprint{APS/123-QED}

\title{Superconducting transmon qubit-resonator quantum battery}% Force line breaks with \\
%\thanks{A footnote to the article title}%

\author{Fu-Quan Dou}
\email{doufq@nwnu.edu.cn}
\affiliation{College of Physics and Electronic Engineering, Northwest Normal University, Lanzhou, 730070, China}
%\date{\today}% It is always \today, today,
             %  but any date may be explicitly specified
\author{Fang-Mei Yang}%
\affiliation{College of Physics and Electronic Engineering, Northwest Normal University, Lanzhou, 730070, China}

\begin{abstract}
Quantum battery (QB) is the miniature energy storage and release device and plays a crucial role in future quantum technology. Here, an implementation scheme of a QB is proposed on a superconducting circuit which is composed by $N$ coupled transmon qubits and a one-dimensional transmission line resonator. We derive the Hamiltonian of the QB system and investigate its charging performance by considering three decay channels. We find that the presence of the decay channels suppresses the high oscillation of the energy storage process, thereby realizing a stable and powerful QB. In particular, compared with the resonator decay and the qubit relaxation, the qubit dephasing shows a counterintuitive advantage in our QB. We show that the nearest neighbor interaction always have a positive impact on the stable energy and the coupling only significantly influences the maximum charging power in the fully nondegenerate ground state region. We also demonstrate the feasibility of our approach by evaluating the QB performance under experimental parameters.
\end{abstract}
%\keywords{Suggested keywords}%Use showkeys class option if keyword or supplies energy Our results provide a practical approach for the realization of the optimal quantum batteries in future experiments.
                              %display desired
\maketitle

%\tableofcontents
\section{Introduction}
Since Robert Alicki and Mark Fannes proposed the concept of quantum batteries (QBs) in $2013$ \cite{PhysRevE.87.042123}, an immense amount of effort has been paid to obtain QBs that have ultrasmall size, ultralarge capacity, ultrafast charging and ultraslow aging \cite{PhysRevLett.120.117702,PhysRevLett.125.236402,ito2020collectively,PhysRevB.102.245407,PhysRevA.100.043833,PhysRevA.102.060201,wang2021fluctuations,PhysRevA.104.042209}. Compared to chemical batteries, which convert chemical energy into electric energy through reactions between two species with different chemical properties \cite{Li2018}, QBs are constituted of quantum systems, exploiting quantum resources (such as quantum coherence and quantum entanglement) to store and transfer energy \cite{campaioli2018quantum,PhysRevB.99.205437,Cruz2022}.

The QB has been extensively studied and most theoretical progresses have been achieved recently, ranging from constructing QB models \cite{Rosa2020,Moraes2021,Dou2020,Dou2021,dou2022charging,PhysRevE.101.062114,PhysRevA.104.032606,PhysRevA.106.032212,PhysRevResearch.4.013172,PhysRevLett.120.117702,PhysRevA.97.022106,PhysRevB.98.205423,PhysRevB.100.115142,PhysRevA.103.052220,PhysRevE.100.032107,Alicki2019,ito2020collectively,PhysRevResearch.2.023095,PhysRevLett.125.236402,PhysRevB.102.245407,Chen2020,PhysRevA.106.022618,landi2021battery,Shaghaghi2022}, analyzing roles of quantum resources \cite{PhysRevLett.111.240401,PhysRevLett.118.150601,PhysRevE.102.052109,PhysRevA.104.L030402,PhysRevB.104.245418,imai2022work,PhysRevLett.128.140501,PhysRevLett.122.047702,Gumberidze2019,centrone2021charging} and many-body interactions \cite{zhang2018enhanced,PhysRevA.104.032606,PhysRevA.97.022106,PhysRevA.106.032212,PhysRevResearch.4.013172,PhysRevE.99.052106,PhysRevA.101.032115,PhysRevE.104.024129,salvia2022quantum,PhysRevB.105.115405,PhysRevA.105.L010201,PhysRevA.105.022628,barra2022efficiency}, researching effects of environment \cite{Liu2019,Carrega2020,PhysRevE.104.064143,PhysRevA.104.032207,PhysRevE.105.064119,Kamin2020,PhysRevA.102.052223,PhysRevE.104.044116,PhysRevApplied.14.024092,PhysRevA.103.033715,PhysRevA.104.043706,PhysRevB.99.035421,PhysRevE.103.042118,PhysRevE.105.054115,PhysRevLett.122.210601,PhysRevA.102.052223}
and initial state \cite{delmonte2021characterization,landi2021battery,LipkaBartosik2021secondlawof,PhysRevA.104.043706,PhysRevA.105.023718}, to discussing possible implementation schemes \cite{PhysRevLett.124.130601,Cruz2022,quach2020,Hu2022,PhysRevA.106.042601,wenniger2022coherence,gemme2022ibm,Zheng2022}.

Experimentally, nowadays QBs can be promisingly implemented on many different physical platforms \cite{quach2020,Hu2022,PhysRevA.106.042601,wenniger2022coherence,gemme2022ibm,Zheng2022}. The first experimental realization of the QB has been reported by using an organic semiconductor as an ensemble of two-level systems coupled to a microcavity \cite{quach2020}. Another experiment characterizing a QB has been realized with a semiconductor quantum dot embeded in an optical microcavity, and the charging and discharging process of the QB are featured by the energy exchanges between the solid-state qubit and light fields \cite{wenniger2022coherence}. Besides, the superconducting circuit is also one of the experimental platforms for realizing QBs \cite{PhysRevE.100.032107,Dou2021,PhysRevResearch.4.033216,Hu2022,Zheng2022}. In contrast to other platforms, superconducting circuits can be artificially designed and fabricated for different research purposes. Their energy levels and the coupling strength between superconducting circuits and their electromagnetic environments can be adjusted by external parameters \cite{You2005,H.2013,RevModPhys.85.623,Gu2017,PRXQuantum.2.040202}. Recently several experimental realizations of QBs based on superconducting circuits have also been reported including the quantum phase battery \cite{Strambini2020}, the transmon qutrit QB \cite{Hu2022} and the Xmon qutrit QB \cite{Zheng2022}. The quantum phase battery consists of an n-doped InAs nanowire with unpaired-spin surface states based on a hybrid superconducting circuit. The charging and discharging process is achieved as continuous tuning of phase bias by an external in-plane magnetic field \cite{Strambini2020}. The transmon qutrit QB is made out of a transmon three-level system coupled to an external field. Its stable charging process is achieved utilizing the stimulated Raman adiabatic passage to bypass unwanted spontaneous discharge or attenuation \cite{Hu2022}. The Xmon qutrit QB, designed on a superconducting Xmon qutrit, is charged via the external driving fields. The stable energy storage process is explored by a freezing phenomenon of populations, which is observed by building a shortcut to adiabaticity in the three-level open system and controlling the Markovian dynamics of the open system \cite{Zheng2022}.

Most of works on QBs focussed on the performance of QBs in closed systems \cite{Crescente2020,PhysRevB.98.205423,PhysRevB.102.245407,PhysRevLett.120.117702,PhysRevA.97.022106,PhysRevA.103.052220,PhysRevB.100.115142}, leaving marginal discussions on potential effects due to the unavoidable interaction with environment. In fact, in physical platforms, especially in superconducting circuits, the resonator decay and the qubit decoherence, which can affect the performance \cite{PhysRevA.100.043833,Liu2019,Carrega2020,PhysRevE.104.064143,PhysRevA.104.032207,PhysRevE.105.064119,Kamin2020,PhysRevA.102.052223} or stabilize the energy storage process of QBs \cite{PhysRevE.104.044116,PhysRevApplied.14.024092,PhysRevA.103.033715,PhysRevA.104.043706}, are important influencing factors lying on the way to experimentally implement QBs.

In this work, we propose an implementation scheme of a QB on superconducting circuits platform. Firstly, we present a superconducting circuit composed by $N$ coupled transmon qubits and a one-dimensional (1D) transmission line resonator and derive its Hamiltonian. Secondly, we consider the QB system of $N$ coupled transmon qubits and the 1D transmission line resonator as a charger. The QB's stable and powerful charging process is discussed with three decay channels, i.e., the resonator
decay, the qubit relaxation and the qubit dephasing. Finally, we demonstrate the feasibility of our implementation scheme by evaluating the performance of our QB under experimental parameters.

The paper is organized as follows. In Sec. \ref{section2} we present a superconducting circuit and derive its Hamiltonian. Then we define a QB based on this superconducting circuit and discuss the QB's stable and powerful charging process with three decay channels in Sec. \ref{section3}. The performance of the QB under experimental parameters is evaluated in Sec. \ref{section4}. Finally, a brief discussion and summary are given in Sec. \ref{section5}.
\section{Circuit quantum electrodynamics} \label{section2}
\begin{figure}[htbp]
 \centering
 \includegraphics[width=0.465\textwidth]{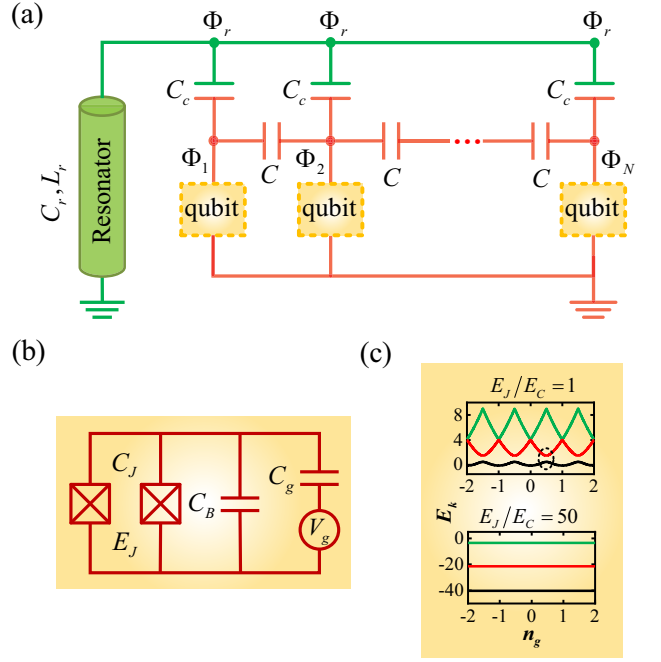}
 \caption{(a) The schematic of a superconducting circuit system composed by $N$ capacitively coupled transmon qubits and a 1D transmission line resonator. (b) Effective circuit diagram of the transmon qubit. (c) Sketch of the energy levels of a transmon qubit as a function of the dimensionless gate charge $n_{g}$ and for different $E_{J}/E_{C}$. Similar to natural atoms, transmon qubit has discrete energy levels, but in contrast to natural atoms, its energy levels can be adjusted by external parameters, e.g., currents and voltages, or magnetic and electric fields \cite{Gu2017}.}
\label{fig.1}
\end{figure}

We consider a superconducting circuit as depicted in Fig. \ref{fig.1}, where each transmon qubit is coupled to its nearest neighbor qubit via capacitance $C$ and coupled to a 1D transmission line resonator with capacitance $C_{r}$ and inductance $L_{r}$ via capacitance $C_{c}$. Each transmon qubit consists two Josephson junctions with capacitance $C_{J}$ and Josephson energy $E_{J}$. The two Josephson junctions are shunted by an additional large capacitance $C_{B}$ and coupled to the gate electrode $V_{g}$ through capacitance $C_{g}$. The whole circuit is described by the following Lagrangian
\begin{eqnarray}
\begin{aligned}
\mathcal{L}&=\mathcal{L}_{r}+\sum_{i=1}^{N}\left[\mathcal{L}_{qi}+\frac{1}{2}C_{c}(\dot{\Phi}_{r}-\dot{\Phi}_{i})^{2}\right]\\
&+\sum_{i=1}^{N-1}\frac{1}{2}C(\dot{\Phi}_{i}-\dot{\Phi}_{i+1})^{2},
\end{aligned}
\end{eqnarray}
where $\Phi_{k} (k=r,i$ and $i=1,...,N)$ is the flux at each node, its derivative to time $\dot{\Phi}_{k}$ represents the voltage at each node, $\mathcal{L}_{r}$ is the Lagrangian for the transmission line resonator, $\mathcal{L}_{qi}$ is the Lagrangian for the $i$th transmon qubit, which we can write as
\begin{eqnarray}
\begin{aligned}
\mathcal{L}_{qi}&=C_{J}\dot{\Phi}_{i}^{2}+\frac{1}{2}C_{B}\dot{\Phi}_{i}^{2}\\
&+\frac{1}{2}C_{g}(V_{g}-\dot{\Phi}_{i})^{2}+2E_{J}cos\delta.
\end{aligned}
\end{eqnarray}
Here $\delta=2\pi\Phi_{i}/\Phi_{0}$ is the gauge-invariant phase difference between the superconductors, $\Phi_{0}=\hbar/2e$ is the flux quantum. For simplicity, we define the abbreviation $C_{t}=C_{0}+2C, C_{0}=2C_{J}+C_{B}+C_{g}+C_{c}$ and then the Lagrangian becomes
\begin{eqnarray}
\begin{aligned}
\mathcal{L}&=\mathcal{L}_{r}-\sum_{i=1}^{N}C_{c}\dot{\Phi}_{r}\dot{\Phi}_{i}\\
&+\sum_{i=1}^{N}(\frac{1}{2}C_{t}\dot{\Phi}_{i}^{2}-C_{g}V_{g}\dot{\Phi}_{i}+2E_{J}cos\delta)\\
&-\frac{1}{2}C(\dot{\Phi}_{1}^{2}+\dot{\Phi}_{N}^{2})-\sum_{i=1}^{N-1}C\dot{\Phi}_{i}\dot{\Phi}_{i+1}.
\end{aligned}
\end{eqnarray}

Performing a Legendre transforming via the relation $H=\sum_{k}Q_{k}\dot{\Phi}_{k}-\mathcal{L}$ and $Q_{k}=\partial\mathcal{L}/\partial\dot{\Phi}_{k}$, we can write the classical Hamiltonian as
\begin{eqnarray}
\begin{aligned}
\label{eq4}
H&=H_{r}-\sum_{i=1}^{N}C_{c}\dot{\Phi}_{r}\dot{\Phi}_{i}+\frac{1}{2}\vec{\dot{\Phi}}\mathcal{C}\vec{\dot{\Phi}}^{T}-\sum_{i=1}^{N}2E_{J}cos\delta,
\end{aligned}
\end{eqnarray}
where $\vec{\dot{\Phi}}=(\dot{\Phi}_{1},\dot{\Phi}_{2},...,\dot{\Phi}_{N})$ is the vector of voltage and $\mathcal{C}$ is the capacitance matrix defined
\begin{equation*}
\mathcal{C}=
	\begin{bmatrix}
    C_{0}+C&-C& & & &\\
	-C&C_{0}+2C&-C& &\\
	 &-C&C_{0}+2C&-C& &\\
	 & &-C&\ddots&\ddots&\\
     & & &\ddots& &
	\end{bmatrix}.
\end{equation*}

The charge $Q_{i}$ is conjugate to the flux $\Phi_{i}$ which obeys the commutation relation $[\Phi_{i},Q_{j}]=i\hbar\delta_{ij}$. By following the usual quantization procedure of the qubit \cite{PhysRevLett.129.087001,PhysRevA.94.033850,PhysRevLett.127.237702} we obtain
\begin{eqnarray}
\begin{aligned}
\vec{\dot{\Phi}}^{T}=2e\mathcal{C}^{-1}(\vec{n}-n_{g})^{T}
\end{aligned}
\end{eqnarray}
where $\vec{n}=(n_{1},n_{2},...,n_{N})$ is the vector of the Cooper pair number, $n_{i}=Q_{i}/2e$ is the number of Cooper pairs transferred between the islands and $n_{g}=-C_{g}V_{g}/2e$  is the dimensionless gate charge.

By quantizing the transmission line resonator mode, we express $\dot{\Phi}_{r}=\sqrt{\hbar\omega_{r}/(C_{r}+NC_{c})}(a+a^{\dag}), H_{r}=\hbar\omega_{r}a^{\dag}a$ with $a (a^{\dag})$ being the annihilation (creation) operator of the resonator \cite{PhysRevA.69.062320,RevModPhys.93.025005}. Note that here we only consider the lowest resonant mode of the transmission line resonator. Using these relations in Eq. (\ref{eq4}), the quantized Hamiltonian
\begin{eqnarray}
\begin{aligned}
\label{eq6}
H&=\hbar\omega_{r}a^{\dag}a-\frac{2eC_{c}}{C_{0}}\sqrt{\frac{\hbar\omega_{r}}{C_{r}+NC_{c}}}\sum_{i=1}^{N}(a+a^{\dag})(n_{i}-n_{g})\\
&+\sum_{i=1}^{N}\left[2e^{2}\mathcal{C}_{ii}^{-1}(n_{i}-n_{g})^{2}-2E_{J}cos\delta\right]\\
&+\sum_{i<j}^{N}4e^{2}\mathcal{C}_{ij}^{-1}(n_{i}-n_{g})(n_{j}-n_{g}).
\end{aligned}
\end{eqnarray}

It is useful to introduce the annihilation (creation) operator $b_{i} (b_{i}^{\dagger})$ of the $i$th transmon and the quantized Hamiltonian takes the form (hereafter we set $\hbar=1$)
\begin{eqnarray}
\begin{aligned}
H&=\omega_{r}a^{\dag}a+I\sqrt{\frac{\omega_{q}\omega_{r}E_{C}C_{c}^2}{e^{2}(C_{r}+NC_{c})}}\sum_{i=1}^{N}(a+a^{\dag})(b_{i}-b_{i}^{\dagger})\\
&+\omega_{q}\sum_{i=1}^{N}b_{i}^{\dagger}b_{i}-\frac{\omega_{q}}{2}\sum_{i<j}\beta^{|i-j|}(b_{i}-b_{i}^{\dagger})(b_{j}-b_{j}^{\dagger}),
\end{aligned}
\end{eqnarray}
where $E_{C}=e^{2}/2C_{0}, \delta=2\sqrt{E_{C}/\omega_{q}}(b_{i}+b_{i}^{\dagger}), n_{i}-n_{g}=-I\sqrt{\omega_{q}/E_{C}}(b_{i}-b_{i}^{\dagger})/4, \beta=C/(C_{0}+C)$, $I$ represents imaginary unit and $\omega_{q}=\sqrt{16E_{C}E_{J}}$ is the frequency of the transmon.

In a regime where the anharmonicity of the transmon \cite{PhysRevA.76.042319} is large compared to the detuning between the transmon and the resonator, one can reduce the transmon to a two-level system \cite{PhysRevA.105.022423,PhysRevA.104.053509,PhysRevA.87.062325,PhysRevB.77.180502,Houck2009} (the specific analysis is shown in the Appendix \ref{appendix1}). The Hamiltonian can be truncated to the two lowest transmon levels ($|0\rangle$ and $|1\rangle$) and written as
\begin{eqnarray}
\begin{aligned}
H&=\omega_{r}a^{\dag}a+I\sqrt{\frac{\omega_{q}\omega_{r}E_{C}C_{c}^2}{e^{2}(C_{r}+NC_{c})}}\sum_{i=1}^{N}(a+a^{\dag})(\sigma^{-}_{i}-\sigma^{+}_{i})\\
&-\frac{\omega_{q}}{2}\sum_{i=1}^{N}\sigma^{z}_{i}+\frac{I^{2}\omega_{q}}{2}\sum_{i<j}\beta^{|i-j|}(\sigma^{-}_{i}-\sigma^{+}_{i})(\sigma^{-}_{j}-\sigma^{+}_{j}),\\
&=\omega_{r}a^{\dag}a+\sqrt{\frac{\omega_{q}\omega_{r}E_{C}C_{c}^2}{e^{2}(C_{r}+NC_{c})}}\sum_{i=1}^{N}(a+a^{\dag})\sigma^{y}_{i}\\
&-\frac{\omega_{q}}{2}\sum_{i=1}^{N}\sigma^{z}_{i}+\frac{\omega_{q}}{2}\sum_{i<j}\beta^{|i-j|}\sigma^{y}_{i}\sigma^{y}_{j},
\end{aligned}
\end{eqnarray}
where $\sigma^{y}_{i}=I(\sigma^{-}_{i}-\sigma^{+}_{i}), \sigma^{z}_{i}=|0\rangle\langle0|-|1\rangle\langle1|, \sigma^{-}_{i}=|0\rangle\langle1|$ and $\sigma^{+}_{i}=|1\rangle\langle0|$ are the Pauli operators describing the $i$th transmon qubit. We change $\sigma_{i}^{y}\rightarrow\sigma_{i}^{x},\sigma_{i}^{z}\rightarrow-\sigma_{i}^{z}$ in a rotating frame and ignore the long-range interaction which is small enough compared with the nearest neighbor interaction between qubits. The final quantized Hamiltonian
\begin{eqnarray}
\begin{aligned}
\label{eq9}
&H=H_{r}+H_{q}+H_{r-q},
\end{aligned}
\end{eqnarray}
where
\begin{eqnarray}
\begin{aligned}
\label{eq10}
&H_{r}=\omega_{r}a^{\dag}a,\\
&H_{q}=\frac{\omega_{q}}{2}\sum_{i=1}^{N}\sigma_{i}^{z}+J\sum_{i=1}^{N-1}\sigma^{x}_{i}\sigma^{x}_{j},\\
&H_{r-q}=g\sum_{i=1}^{N}(a+a^{\dag})\sigma^{x}_{i}.
\end{aligned}
\end{eqnarray}
Here, $H_{r}$ is the Hamiltonian of the transmission line resonator, $\omega_{r}=2\pi/\sqrt{L_{r}(C_{r}+NC_{c})}$ is the frequency of the resonator, $H_{q}$ is the Hamiltonian of the transmon qubits, $J=\omega_{q}\beta/2$ is the nearest neighbor interaction strength between the qubits, $H_{r-q}$ is the coupling Hamiltonian between the resonator and the qubits, and $g=\sqrt{\omega_{q}\omega_{r}E_{C}C_{c}^2/(e^{2}(C_{r}+NC_{c}))}$ is the coupling strength between the qubits and the resonator.

Previous works \cite{PhysRevE.100.032107,Dou2021,PhysRevResearch.4.033216,Hu2022,Zheng2022} of QBs have found that the superconducting circuit is one of the experimental platforms for realizing QBs. Next we will define a QB model based on this superconducting circuit and discuss its charging process.

\section{The QB model and its charging process with three decay channels} \label{section3}
The total Hamiltonian of the QB system can be written as Eq. (\ref{eq9}), where a 1D transmission line resonator $H_{r}$ develops the role of a quantum charger and $N$ capacitively coupled transmon qubits $H_{q}$ as a QB. The presence of transmon qubits induces a strong change in the impedance of the circuit through which microwave photons propagate, enabling qubit-photon interactions \cite{RevModPhys.91.025005}. In our model the interaction $H_{r-q}$ is capacitive and play a key role in charging process of our QB. %can be used to charge our QB.

Similar to other quantum systems, the QB system can be considered as an open system because of the unavoidable interaction with environment. In order to provide a more realistic description, we consider the effects of the resonator decay and the qubit decoherence (mainly resulting from relaxation and dephasing) during the QB's charging process. The charging dynamics of the QB is obtained by solving the quantum master equation
\begin{equation}
\dot{\rho}(t)=I[\rho(t),H]+\kappa\mathbb{L}[a]+\Gamma_{1}\mathbb{L}[J_{-}]+\Gamma_{2}\mathbb{L}[J_{z}],
\end{equation}
where $J_{-}=\sum_{i=1}^{N}\sigma_{i}^{-}, J_{z}=\sum_{i=1}^{N}\sigma_{i}^{z}$. $\rho(t)=|\psi(t)\rangle\langle \psi(t)|$ is the density matrix of the considered system, $\mathbb{L}[A]=A\rho A^{\dag}-(A^{\dag}A\rho+\rho AA^{\dag})/2$ is the Lindblad operator. There are three decay channels corresponding to the resonator decay rate $\kappa$, the individual qubit relaxation rate $\Gamma_{1}$ and dephasing rate $\Gamma_{2}$, respectively.

We take the initial state of the QB in its ground state $|G\rangle$ and the initial state of the resonator in the Fock state $|n_{r}\rangle$ with the full energy. Thus, the initial state of the whole system is as follows
\begin{equation}
|\psi(0)\rangle=|G\rangle\otimes|n_{r}\rangle.
\end{equation}

The average of $S_{z}$ can be represented as
\begin{equation}
\langle S_{z}\rangle=Tr[S_{z}\rho_{q}(t)],
\end{equation}
where $S_{z}=\sum_{i=1}^{N}\sigma_{i}^{z}$ and $\rho_{q}(t)=Tr_{r}[\rho(t)]$ is the reduced density matrix of the QB. The average at the QB's initial state can be expressed as $\langle S_{z}\rangle_{G}$. $\langle S_{z}\rangle_{G}/N=p_{1}-p_{0}$ where $p_{0}$ and $p_{1}$ are the populations of the ground state $|0\rangle$ and excited state $|1\rangle$ of the transmon qubit \cite{Carrega2020}.

When the coupling strength $g$ between the qubits and the resonator is turned on, the charging process immediately starts the energy exchange between the qubits and the resonator. The energy in the QB at time $t$ is given by
\begin{equation}
E(t)=Tr[H_{q}\rho_{q}(t)],
\end{equation}
The charging energy at time $t$ is the difference in energy between the final and initial states
\begin{equation}
\begin{aligned}
\label{eq13}
&\Delta E(t)=E(t)-E(0),
\end{aligned}
\end{equation}
where $E(0)=E_{G}$ is the ground-state energy of the QB. The average charging power is given by
\begin{equation}
\begin{aligned}
\label{eq14}
&P(t)=E(t)/t.
\end{aligned}
\end{equation}

The charging energy $\Delta E(t)$, the average charging power $P(t)$, the stable energy $E_s=\Delta E(\infty)$ and the maximum charging power $P_{max}=max[P(t)]$ are used to characterize the performance of the QB with decay channels.

\begin{figure}[htbp]
 \centering
 \includegraphics[width=0.5\textwidth]{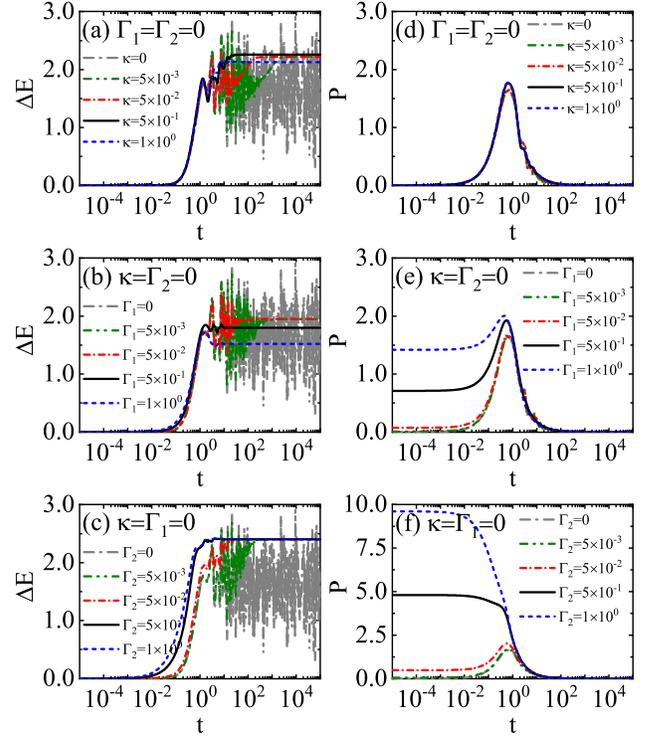}
 \caption{(a)-(c) The time evolution of the charging energy $\Delta E(t)$ (in unit of $\hbar\omega$) and (d)-(f) the average charging power $P(t)$ (in unit of $\hbar\omega^{2}$). The gray dash-dotted curve represents the case of closed system, other curves represent the cases of open system. We set $g=J=1$.}
\label{fig2}
\end{figure}
In the following, we focus on the resonance regime ($\omega_{q}=\omega_{r}=\omega$ in Eq. (\ref{eq9})) unless specifically noted. For simplicity, we treat all parameters are in unit of $\omega$ and set $N=n_{r}=3, \omega=1$. Figures \ref{fig2}(a)-(c) illustrate the time evolution of the charging energy in different cases. In a closed system, corresponding to $\kappa=\Gamma_{1}=\Gamma_{2}=0$, the charging energy is unstable and is highly oscillatory due to the existence of counter-rotating terms in the system.
In open systems, the quantum interference, caused by the interplay of collective effects and three decay channels (the resonator decay, the qubit relaxation and the qubit dephasing), suppresses this highly oscillatory phenomenon and can lead to steady state of the QB. As the steady state is decoupled from the environment, the charging energy of the QB is stable while the charger is present, even in the open system \cite{PhysRevApplied.14.024092,PhysRevA.104.032207}. In this case, the absorption and radiation of the photons in the QB reach a dynamic balance. In particular, as long as the coupling strength $g$ between the qubits and the resonator is turned on, the charging process immediately starts but there is little charging energy. The time evolution of the average charging power $P(t)$ is shown in Fig. \ref{fig2}(d)-(f). When considering the qubit relaxation or dephasing, at the start of the energy storage the average charging power of the QB remarkably increases as the qubit relaxation and dephasing rate increase, which indicates faster charging process compared with the closed system.

\begin{figure}[htbp]
 \centering
 \includegraphics[width=0.5\textwidth]{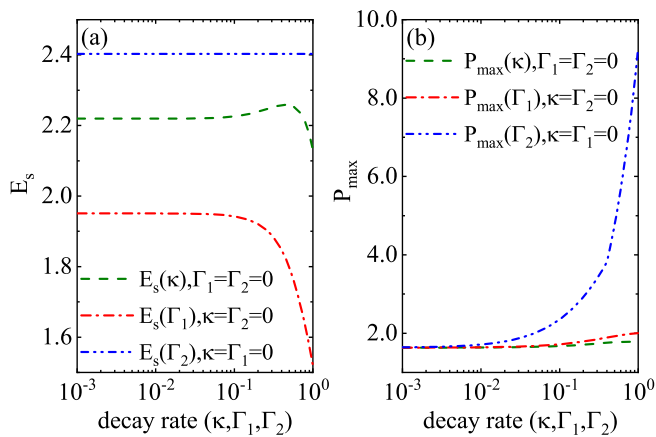}
 \caption{The stable energy $E_s$ (in unit of $\hbar\omega$) and the maximum charging power $P_{max}$ (in unit of $\hbar\omega^{2}$) as a function of the decay channels (the resonator decay, the qubit relaxation and the qubit dephasing). We set $g=J=1$.}
\label{fig.3}
\end{figure}
\begin{figure}[htbp]
 \centering
 \includegraphics[width=0.5\textwidth]{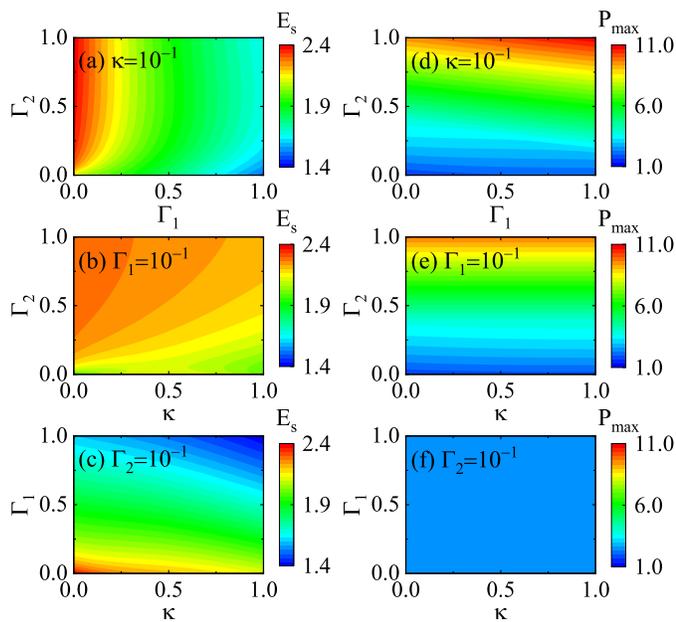}
 \caption{Contour plot of the stable energy $E_s$ (in unit of $\hbar\omega$) and the maximum charging power $P_{max}$ (in unit of $\hbar\omega^{2}$) as a function of the decay channels (the resonator decay, the qubit relaxation and the qubit dephasing). We set $g=J=1$.}
\label{fig.4}
\end{figure}
The effects of the decay channels on the stable energy $E_s$ and the maximum charging power $P_{max}$ are shown in Fig. \ref{fig.3} and Fig. \ref{fig.4}. We observe several phenomenons as follow: (\romannumeral1) Consistent with the analysis results in Ref. \cite{quach2020}, the qubit dephasing plays a crucial role in the stable and powerful charging process of our QB. The qubit dephasing, a combination of pure dephasing and energy relaxation, causes the loss of coherence of a quantum state \cite{PhysRevA.76.042319,Krantz2019}. The pure dephasing part, usually being treated within the adiabatic approximation, can modify the transition frequency of the qubit \cite{PhysRevA.76.042319}. However, the energy relaxation part breaks the adiabatic approximation and induces transitions between the qubit states. This suppresses the qubit energy decay into the environment, so that we can obtain the QB with high stable energy and short charging time \cite{quach2020}, as shown by the blue line in Fig. \ref{fig.3} and Fig. \ref{fig.4}(d)-(f). (\romannumeral2) The qubit relaxation plays inhibition role in the stable charging process of our QB. The qubit relaxation, which makes transitions from the excited state $|e\rangle$ to the ground state $|g\rangle$ of the qubits and suppresses the population inversion of the quantum states \cite{PhysRevA.76.042319,Krantz2019}, leads to a remarkable decay on the stable energy of our QB, see the red line in Fig. \ref{fig.3}(a) and Fig. \ref{fig.4}(a)-(c). (\romannumeral3) The resonator decay causes photons in the resonator to flow into the environment. In a Tavis-Cummings QB \cite{PhysRevA.104.043706}, as photons flow into the environment, the absorption and radiation of photons no longer remain in balance. Finally, the charging energy of the Tavis-Cummings QB will be exhausted with the photons of the resonator completely leaked \cite{PhysRevA.104.043706}. However, compared with the Tavis-Cummings QB, an external coherent driving field in Ref. \cite{PhysRevA.103.033715} continuously provides photons to the resonator with decay, resulting in a stable charging process. In our QB system, the counter-rotating terms and collective effects play the same role as the coherent field, causing the number of photons to reach a dynamic balance when considering the resonator decay. The stable energy and the maximum charging power are almost unchanged as the resonator decay rate increases in Fig. \ref{fig.3} (the olive green line) and Fig. \ref{fig.4}(a)-(c).

\begin{figure}[htbp]
 \centering
 \includegraphics[width=0.5\textwidth]{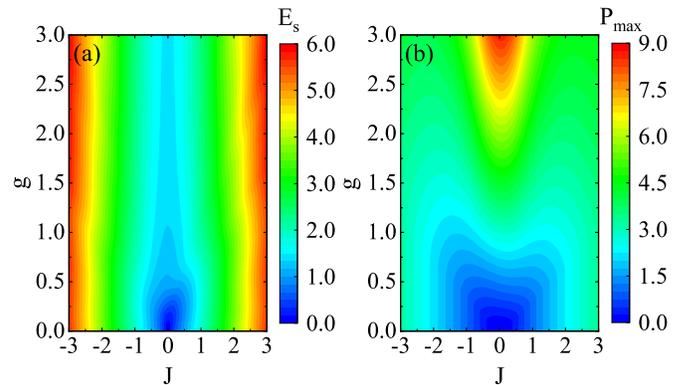}
 \caption{Contour plot of the stable energy $E_s$ (in unit of $\hbar\omega$) and the maximum charging power $P_{max}$ (in unit of $\hbar\omega^{2}$) as a function of the nearest neighbor interaction strength $J$ between the qubits and the coupling strength $g$ between the qubits and the resonator. We set $\kappa=\Gamma_{1}=\Gamma_{2}=10^{-1}$.}
\label{fig.5}
\end{figure}
\begin{figure}[htbp]
 \centering
 \includegraphics[width=0.5\textwidth]{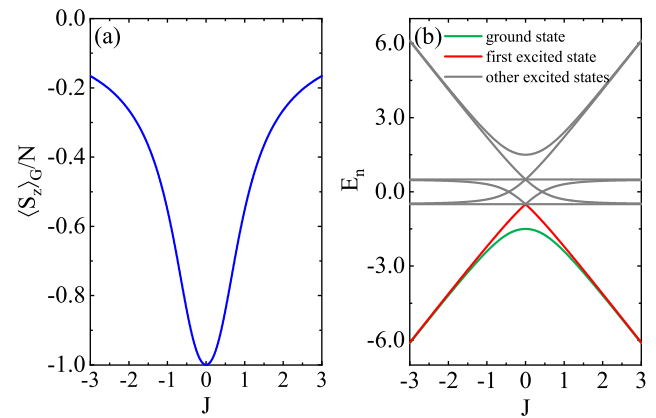}
 \caption{(a) The population difference and (b) the energy spectrum dependent on the nearest neighbor interaction strength $J$. The blue line represent the population difference of the ground state and excited state of each qubit at the QB's initial state. The red line, green line and gray lines represent the energy $E_{n}$ for the ground state, first excited state and other excited states of the QB system, respectively.}
\label{fig.6}
\end{figure}
\renewcommand{\tabcolsep}{0.07cm}
\renewcommand{\arraystretch}{1.5}
\begin{table*}[htbp]
  \centering
  \caption{The stable energy $E_s$ and the maximum charging power $P_{max}$ of the QB under experimental parameters.}
      \begin{tabular}{ccccccccccccc}
      \hline\hline
      \textit{Reference}  &$\omega_{r}/2\pi$  &$\omega_{q}/2\pi$  &$g/2\pi$  &$\kappa/2\pi$  &$\Gamma_{1}/2\pi$  &$\Gamma_{2}/2\pi$  &$g/\omega_{r}$ (coupling  &charging &$J/2\pi$   &$E_s$  &$P_{max}$\\
      \textit{} &$(GHz)$  &$(GHz)$  &$(MHz)$  &$(MHz)$  &$(MHz)$  &$(MHz)$  &regime)  &scheme  &$(MHz)$  &$(neV)$  &$(eV/s)$
      \\ \hline
      \multirow{2}{*}{\textit{Ref. \cite{Schuster2007}}} &\multirow{2}{*}{5.7}  &\multirow{2}{*}{6.9}  &\multirow{2}{*}{105}  &\multirow{2}{*}{0.25}  &\multirow{2}{*}{1.8}  &\multirow{2}{*}{1.0}  &\multirow{2}{*}{0.020 (SC)}   &parallel &0 &$2.2574\times10^{1}$  &$7.0347\times10^{3}$\\ & & & & & & & &collective &105 &$2.4003\times10^{2}$  &$1.8178\times10^{4}$\\
      \multirow{2}{*}{\textit{Ref. \cite{Bishop2009}}}  &\multirow{2}{*}{6.92}  &\multirow{2}{*}{6.92}  &\multirow{2}{*}{173.5}  &\multirow{2}{*}{0.30}  &\multirow{2}{*}{0.094}  &\multirow{2}{*}{0.227}  &\multirow{2}{*}{0.025 (SC)}   &parallel  &0  &$1.9964\times10^{2}$  &$6.7230\times10^{4}$\\ & & & & & & & &collective &173.5 &$6.0026\times10^{2}$  &$9.4678\times10^{4}$\\
      \multirow{2}{*}{\textit{Ref. \cite{PhysRevB.95.224515}}} &\multirow{2}{*}{6.23}  &\multirow{2}{*}{3.586}  &\multirow{2}{*}{455}  &\multirow{2}{*}{29.3}  &\multirow{2}{*}{38}  &\multirow{2}{*}{0}  &\multirow{2}{*}{0.071 (SC)} &parallel  &0  &$1.7015\times10^{2}$  &$2.9198\times10^{4}$\\  & & & & & & & &collective &455 &$3.3901\times10^{2}$  &$8.2112\times10^{4}$\\
      \multirow{2}{*}{\textit{Ref. \cite{Bosman2017}}} &\multirow{2}{*}{4.603}  &\multirow{2}{*}{10.67}  &\multirow{2}{*}{897}  &\multirow{2}{*}{3}  &\multirow{2}{*}{20}  &\multirow{2}{*}{0}  &\multirow{2}{*}{0.195 (USC)}   &parallel  &0 &$5.7195\times10^{2}$  &$1.5266\times10^{5}$\\  & & & & & & &  &collective &897 &$1.1007\times10^{3}$  &$3.4661\times10^{5}$\\
      \multirow{2}{*}{\textit{ideal case1}} &\multirow{2}{*}{5}  &\multirow{2}{*}{10}  &\multirow{2}{*}{2500}  &\multirow{2}{*}{1}  &\multirow{2}{*}{1}  &\multirow{2}{*}{1}  &\multirow{2}{*}{0.500 (USC)}  &parallel  &0  &$1.5508\times10^{4}$  &$9.9804\times10^{5}$\\  & & & & & & & &collective  &2500  &$3.2328\times10^{4}$  &$1.5388\times10^{6}$\\
      \multirow{2}{*}{\textit{ideal case2}} &\multirow{2}{*}{5}  &\multirow{2}{*}{10}  &\multirow{2}{*}{5000}  &\multirow{2}{*}{1}  &\multirow{2}{*}{1}  &\multirow{2}{*}{1}  &\multirow{2}{*}{1.000 (DSC)}  &parallel  &0  &$3.5445\times10^{4}$  &$2.5253\times10^{6}$\\  & & & & & & & &collective &5000 &$5.7538\times10^{4}$  &$3.0394\times10^{6}$\\ \hline\hline
      \end{tabular}%
  \label{TAB.1}%
\end{table*}
For a clearer and more comprehensive understanding of the performance of our QB, we further investigate the stable energy $E_s$ and the maximum charging power $P_{max}$ as a function of the nearest neighbor interaction strength $J$ and the coupling strength $g$ shown in Fig. \ref{fig.5}. It is interesting to note that in our model the nearest neighbor interaction (whether repulsive or attractive) always has a positive impact on the stable energy. This is because the initial state of each qubit which makes up our QB system is the superposition state and the population of qubit's excited state always increases with the nearest neighbor interaction increasing, see Fig. \ref{fig.6}(a). However, as the coupling strength increases, the maximum charging power only significantly increases in the fully nondegenerate ground state region, i.e., $-1<J<1$ in the energy spectrum (as shown in Fig. \ref{fig.6}(b)). When the QB is initially prepared in the nearly degenerate ground state regime, the maximum charging power hardly change with the coupling strength due to the nearly gapless energies leading to the breakdown of the adiabatic condition and the formation of non-adiabatic excitations \cite{PhysRevResearch.4.L022017}.
\section{Evaluate the performance of the QB under experimental parameters} \label{section4}
Superconducting circuits are shown to be an excellent platform to study light-matter interactions in the microwave regime of frequencies \cite{RevModPhys.91.025005}. Early studies \cite{PhysRevA.69.062320,Wallraff2004} of qubit-resonator systems have found that a superconducting qubit interacting with a microwave resonator follows cavity quantum electrodynamics. Above we propose a QB composed by $N$ capacitively coupled transmon qubits and a 1D transmission line resonator as charger, with the transmon qubit playing the role of an artificial atom and the 1D transmission line resonator emulating the cavity.

In this section we evaluate the charging performance of our QB in two different situations: the collective charging (the QB with interaction between transmon qubits and qubits are collectively charged by a shared resonator) and the parallel charging (the QB without interaction between the transmon qubits and each qubit is charged by an independent resonator) \cite{PhysRevA.101.032115,PhysRevE.102.052109}. Our simulation is based on Ref. \cite{Schuster2007,Bishop2009,PhysRevB.95.224515,Bosman2017,RevModPhys.91.025005} and we set $ N=n_{r}=3$. The simulation results are shown in Table. \ref{TAB.1}. In the strong coupling regime (SC) and the weak ultrastrong coupling regime (USC), the stable energy $E_s$ and the maximum charging power $P_{max}$ in the collective charging process is significantly higher than that in the parallel charging process. Moreover, when we consider the nearest neighbor interaction in the strong USC and the deep-strong coupling regime (DSC), the stable energy $E_s$ and the maximum charging power $P_{max}$ are greatly optimized, as shown in the ideal cases of Table. \ref{TAB.1}. Experimentally, the strong USC and the DSC have now been observed using superconducting circuits with flux qubits \cite{Niemczyk2010,PhysRevLett.105.237001,PhysRevB.93.214501,PhysRevA.95.053824,Yoshihara2017,PhysRevLett.120.183601,Forn-Daz2017}. Meanwhile, the circuit quantum electrodynamics realized the U/DSC between transmon qubits and the microwave field inside an on-chip transmission-line resonator have been extensively studied in recent years \cite{Wallraff2004,PhysRevA.69.062320,PhysRevLett.103.083601,Langford2017,Braumller2017}. We believe that our QB will be realized with the gradual development of experimental techniques in the future.
\section{Discussions and conclusions} \label{section5}
%\nocite{*}
The decoherence processes of the QB, caused by the inevitable interactions with environment, generally deactivate the QB’s charging-storing-discharging process, which is called aging of the QB \cite{PhysRevA.100.043833,PhysRevA.102.060201}. A good QB should have not only ultrasmall size, ultralarge capacity and ultrafast charging, but also ultraslow aging which requires that the quantum systems constituting the QB have a long lifetimes. In contrast to other quantum systems, the superconducting circuits offer flexibility, tunability, scalability and strong coupling to external fields, but have relatively short lifetimes ($\lesssim1ms$) because of the macroscopicity of circuit design \cite{RevModPhys.85.623}. However, the QB's charging-storing-discharging process can be driven faster than the superconducting qubits lifetimes. Examples of systems relevant to our analysis include the Xmon qutrit QB and the transmon qutrit QB \cite{Zheng2022,Hu2022}. The charging-storing-discharging process of the Xmon qutrit QB takes only $3\mu s$, which is far less than the QB lifetimes of $6-9.5\mu s$ \cite{Zheng2022}. The transmon qutrit QB's stable charging process is achieved within $0.3\mu s$ when the QB operates at $51.4-79.4kHz$ decay rates \cite{Hu2022}. Similarly, we can control the energy storage process of our QB in the range of $10-100ns$, which is much faster than the superconducting qubits lifetimes ($\lesssim1ms$). Under such a condition, the aging of our QB is slow.

Nowadays, quantum technologies are still in their infancy and there is a long way to go before QBs can be implemented in practice. Besides the widely studied powerful charging and stable energy storage processes, there are two challenges also worth mentioning: (\romannumeral1) One challenge is the additional cost of schemes, for example the cost of the preparation process of the initial state \cite{delmonte2021characterization,landi2021battery,LipkaBartosik2021secondlawof,PhysRevA.104.043706,PhysRevA.105.023718,PhysRevA.104.032207}, the switching operation on the charger \cite{Carrega2020} and the sequential measurement process for stabilizing open QBs \cite{PhysRevResearch.2.013095}. (\romannumeral2) Another challenge is the capability to fully transfer the stored energy to consumption centers in a useful way (the skill of extracting useful work) \cite{PhysRevA.102.052223,PhysRevLett.111.240401,Allahverdyan2004}. Some researchers point out that it may find key uses in future fusion power plants, which require large amounts of energy to be charged and discharged in an instant \cite{PhysRevLett.128.140501}. In our scheme, we can construct a quantum charger-battery-load circuit \cite{PhysRevA.106.042601} in order to efficiently extract energy. When the coupling between the QB and the load is open while the coupling between the charger and the QB is closed, the energy in the QB can be extracted to the load by transitioning the QB from the steady state back to its ground state.

In summary, we have proposed an implementation scheme of a QB on superconducting circuits platform and discussed the QB's charging performance with three decay channels (the resonator decay, the qubit relaxation and the qubit dephasing). Our results show that the presence of the decay channels suppresses the high oscillation of energy storage process, thereby realizing a QB with a stable and powerful charging process. Compared with other two decay channels, the qubit dephasing shows a counterintuitive advantage and plays a crucial role in our QB. It induces transitions between the qubit states and suppresses the qubit energy decay into the environment, so that we can obtain a QB with high stable energy and short charging time. We have also investigated how the nearest neighbor interaction strength and the coupling strength affect the QB's charging performance. We have found that the nearest neighbor interaction (whether repulsive or attractive) always have a positive impact on the stable energy because the initial state of each qubit which makes up our QB system is the superposition state. However, as the coupling strength increases, the maximum charging power only significantly increases in the fully nondegenerate ground state region. When the QB is initially prepared in the nearly degenerate ground state regime, the maximum charging power hardly change with the coupling strength due to the nearly gapless energies. Moreover, we have demonstrated that our implementation scheme is feasible by evaluating the performance of our QB under experimental parameters. Our results provide an implementation scheme for realization of the efficient QB in future experiments.
\section*{Acknowledgments}
The work is supported by the National Natural Science Foundation of China (Grant No. 12075193).

\appendix
\section{The behaviors of the transmon in different parameter regimes} \label{appendix1}
The crucial design distinguishing the transmon from the Cooper pair box (CPB) is a shunting connection of the two superconductors via a large capacitance $C_{B}$, accompanied by a significantly increase in the ratio of Josephson energy and charging energy $E_{J}/E_{C}$. The price to pay for this increased $E_{J}/E_{C}$ is the reduced anharmonicity $\alpha$ of the transmon qubit \cite{PhysRevA.76.042319,RevModPhys.93.025005}. The anharmonicity $\alpha$ is an important parameter used to characterize the behaviors of CPB and transmon in different parameter regimes, as shown in Table. \ref{TAB.2}. Here, $\alpha=E_{12}-E_{01}$ is the anharmonicity \cite{PhysRevA.76.042319}, $\Delta=\hbar(\omega_{01}-\omega_{r})$ is the detuning between the superconducting qubit and the resonator, $E_{n_{i}(n_{i}+1)}$ and $\omega_{n_{i}(n_{i}+1)}$ are the transition energy and the transition frequency from $|n_{i}\rangle$ state to $|n_{i}+1\rangle$.
\renewcommand{\tabcolsep}{0.07cm}
\renewcommand{\arraystretch}{1.38}
\begin{table}[htbp]
  \centering
  \caption{The behaviors of CPB and transmon in different parameter regimes.}
      \begin{tabular}{cccc}
      \hline\hline
      \textit{}  &$E_{J}/E_{C}$  &$\alpha$  &behavior  \\\hline
      \textit{CPB} &$\ll1$  &large $\alpha$  &two-level system \cite{PhysRevLett.127.237702,PhysRevA.69.062320,Wallraff2004,PhysRevA.94.033850,Ma2021}  \\
      \multirow{3}{*}{\textit{transmon}} &\multirow{3}{*}{$\gg1$}  &$\alpha\gg|\Delta|$  &two-level system \cite{PhysRevA.105.022423,PhysRevA.104.053509,PhysRevA.87.062325,PhysRevB.77.180502,Houck2009}  \\   & &\multirow{2}{*}{$\alpha\ll|\Delta|$} &weakly anharmonic oscillator \\
      \textit{}  &  &  &(multi-level system) \cite{PhysRevA.94.063861,PhysRevA.86.013814,PhysRevA.88.052330,PhysRevLett.114.010501,Pirkkalainen2013}  \\ \hline\hline
      \end{tabular}%
  \label{TAB.2}%
\end{table}

In the small transmon-resonator detuning regime, i.e. $\alpha\gg|\Delta|$, the transition frequency $\omega_{n_{i}(n_{i}+1)}$ from states $|n_{i}\rangle$ to $|n_{i}+1\rangle$ for any $n_{i}\geqslant1$ is strongly off-resonant from resonator frequency $\omega_{r}$. Therefore the resonator photons are blocked from exciting the transmon to states $|n_{i}\geqslant2\rangle$ and the transmon behaves like a quantum two-level system when it interacts with the resonator \cite{PhysRevA.105.022423,PhysRevA.104.053509,PhysRevA.87.062325,PhysRevB.77.180502,Houck2009}. The anharmonicity $\alpha$ decays only algebraically in $E_{J}/E_{C}$, as compared to the exponential suppression of the sensitivity to charge noise in $E_{J}/E_{C}$ \cite{PhysRevA.76.042319,PhysRevA.87.062325}. Thus, the transmon can be engineered to have a large enough anharmonicity $\alpha$, such as $\alpha/2\pi\sim200-500MHz$ \cite{PhysRevA.87.062325,PhysRevA.101.053802,PhysRevB.77.180502,Houck2009}. In this case, it is reasonable to treat the transmon as a two-level system only involving the ground state $|0\rangle$ and the first excited state $|1\rangle$.

\begin{figure}[htbp]
 \centering
 \includegraphics[width=0.5\textwidth]{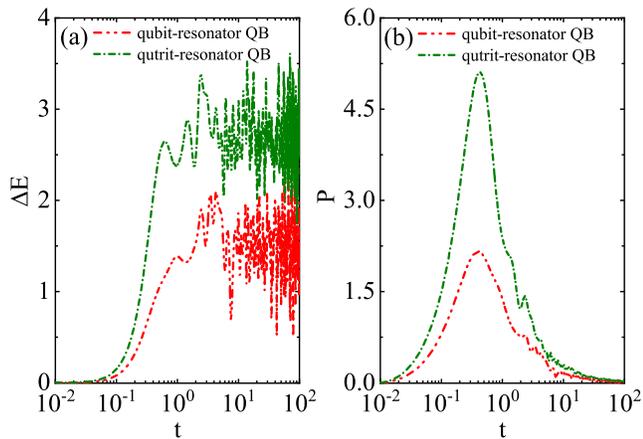}
 \caption{(a) The time evolution of the charging energy $\Delta E(t)$ (in unit of $\hbar\omega$) and (b) the average charging power $P(t)$ (in unit of $\hbar\omega^{2}$). The red curve represents the case of the qubit-resonator QB and the green curve represents the case of qutrit-resonator QB.}
\label{fig.7}
\end{figure}
Differently, in the regime $\alpha\ll|\Delta|$, the transitions between states $|n_{i}+1\rangle$ and $|n_{i}\rangle$ are almost equally likely to be excited by the resonator, causing unwanted transitions to higher excited states. Therefore the transmon behaves almost like a weakly anharmonic oscillator \cite{PhysRevA.86.013814,PhysRevA.88.052330,PhysRevLett.114.010501,Pirkkalainen2013}. In this case, we consider three energy levels of the transmon (including the ground state $|0\rangle$, the first excited state $|1\rangle$ and the second excited state $|2\rangle$) since the main leakage out of the qubit basis comes from the third energy level \cite{PhysRevA.94.063861,PhysRevA.98.052314}. The Hamiltonian of the qutrit-resonator system can be expressed as
\begin{eqnarray}
\begin{aligned}
\label{eqA1}
H'=H'_{r}+H'_{q}+H'_{r-q},
\end{aligned}
\end{eqnarray}
where
\begin{eqnarray}
\begin{aligned}
\label{eqA2}
&H'_{r}=\omega_{r}a^{\dag}a,\\
&H'_{q}=\omega_{q}\sum_{i=1}^{N}S_{i}^{+}S_{i}^{-}+I^{2}J\sum_{i<j}(S_{i}^{-}-S_{i}^{+})(S_{j}^{-}-S_{j}^{+}),\\
&H'_{r-q}=Ig\sum_{i=1}^{N}(a+a^{\dag})(S_{i}^{-}-S_{i}^{+}),
\end{aligned}
\end{eqnarray}
here we treat the annihilation operator of the weakly anharmonic oscillator as $b\equiv|0\rangle\langle 1|+\sqrt{2}|1\rangle\langle 2|+\sqrt{3}|2\rangle\langle 3|+...$ \cite{PhysRevA.91.043846} and let $S^{-}=|0\rangle\langle 1|+\sqrt{2}|1\rangle\langle 2|$ in the qutrit-resonator system. $H'_{r}, H'_{q}, H'_{r-q}$ are the Hamiltonian of the transmission line resonator, the transmon qutrits and the coupling between the resonator and the qutrits, respectively.

Similarly, we define the resonator $H'_{r}$ developing the role of a quantum charger and $N$ capacitively coupled transmon qutrits $H'_{q}$ as a QB. The definitions of the initial state, the charging energy and the average charging power during the charging process are the same as those in Sec. \ref{section3}. For simplicity, we consider a closed system and set $N=3, n_{r}=2N, \omega_{q}=\omega_{r}=\omega=1, g=J=1$. The time evolution of the charging energy $\Delta E(t)$ and the average charging power $P(t)$ are shown in Fig. \ref{fig.7}. We find that compared with the qubit-resonator QB (Eq. \ref{eq9}), the qutrit-resonator QB (Eq. \ref{eqA1}) can store higher energy and charge faster. This phenomenon in Fig. \ref{fig.7} demonstrates that the transition to the second excited state can improve the performance of our QB.

\bibliography{refercence}% Produces the bibliography via BibTeX.

%merlin.mbs apsrev4-1.bst 2010-07-25 4.21a (PWD, AO, DPC) hacked
%Control: key (0)
%Control: author (8) initials jnrlst
%Control: editor formatted (1) identically to author
%Control: production of article title (-1) disabled
%Control: page (0) single
%Control: year (1) truncated
%Control: production of eprint (0) enabled
\begin{thebibliography}{125}%
\makeatletter
\providecommand \@ifxundefined [1]{%
 \@ifx{#1\undefined}
}%
\providecommand \@ifnum [1]{%
 \ifnum #1\expandafter \@firstoftwo
 \else \expandafter \@secondoftwo
 \fi
}%
\providecommand \@ifx [1]{%
 \ifx #1\expandafter \@firstoftwo
 \else \expandafter \@secondoftwo
 \fi
}%
\providecommand \natexlab [1]{#1}%
\providecommand \enquote  [1]{``#1''}%
\providecommand \bibnamefont  [1]{#1}%
\providecommand \bibfnamefont [1]{#1}%
\providecommand \citenamefont [1]{#1}%
\providecommand \href@noop [0]{\@secondoftwo}%
\providecommand \href [0]{\begingroup \@sanitize@url \@href}%
\providecommand \@href[1]{\@@startlink{#1}\@@href}%
\providecommand \@@href[1]{\endgroup#1\@@endlink}%
\providecommand \@sanitize@url [0]{\catcode `\\12\catcode `\$12\catcode
  `\&12\catcode `\#12\catcode `\^12\catcode `\_12\catcode `\%12\relax}%
\providecommand \@@startlink[1]{}%
\providecommand \@@endlink[0]{}%
\providecommand \url  [0]{\begingroup\@sanitize@url \@url }%
\providecommand \@url [1]{\endgroup\@href {#1}{\urlprefix }}%
\providecommand \urlprefix  [0]{URL }%
\providecommand \Eprint [0]{\href }%
\providecommand \doibase [0]{http://dx.doi.org/}%
\providecommand \selectlanguage [0]{\@gobble}%
\providecommand \bibinfo  [0]{\@secondoftwo}%
\providecommand \bibfield  [0]{\@secondoftwo}%
\providecommand \translation [1]{[#1]}%
\providecommand \BibitemOpen [0]{}%
\providecommand \bibitemStop [0]{}%
\providecommand \bibitemNoStop [0]{.\EOS\space}%
\providecommand \EOS [0]{\spacefactor3000\relax}%
\providecommand \BibitemShut  [1]{\csname bibitem#1\endcsname}%
\let\auto@bib@innerbib\@empty
%</preamble>
\bibitem [{\citenamefont {Alicki}\ and\ \citenamefont
  {Fannes}(2013)}]{PhysRevE.87.042123}%
  \BibitemOpen
  \bibfield  {author} {\bibinfo {author} {\bibfnamefont {R.}~\bibnamefont
  {Alicki}}\ and\ \bibinfo {author} {\bibfnamefont {M.}~\bibnamefont
  {Fannes}},\ }\href {\doibase 10.1103/PhysRevE.87.042123} {\bibfield
  {journal} {\bibinfo  {journal} {Phys. Rev. E}\ }\textbf {\bibinfo {volume}
  {87}},\ \bibinfo {pages} {042123} (\bibinfo {year} {2013})}\BibitemShut
  {NoStop}%
\bibitem [{\citenamefont {Ferraro}\ \emph {et~al.}(2018)\citenamefont
  {Ferraro}, \citenamefont {Campisi}, \citenamefont {Andolina}, \citenamefont
  {Pellegrini},\ and\ \citenamefont {Polini}}]{PhysRevLett.120.117702}%
  \BibitemOpen
  \bibfield  {author} {\bibinfo {author} {\bibfnamefont {D.}~\bibnamefont
  {Ferraro}}, \bibinfo {author} {\bibfnamefont {M.}~\bibnamefont {Campisi}},
  \bibinfo {author} {\bibfnamefont {G.~M.}\ \bibnamefont {Andolina}}, \bibinfo
  {author} {\bibfnamefont {V.}~\bibnamefont {Pellegrini}}, \ and\ \bibinfo
  {author} {\bibfnamefont {M.}~\bibnamefont {Polini}},\ }\href {\doibase
  10.1103/PhysRevLett.120.117702} {\bibfield  {journal} {\bibinfo  {journal}
  {Phys. Rev. Lett.}\ }\textbf {\bibinfo {volume} {120}},\ \bibinfo {pages}
  {117702} (\bibinfo {year} {2018})}\BibitemShut {NoStop}%
\bibitem [{\citenamefont {Rossini}\ \emph {et~al.}(2020)\citenamefont
  {Rossini}, \citenamefont {Andolina}, \citenamefont {Rosa}, \citenamefont
  {Carrega},\ and\ \citenamefont {Polini}}]{PhysRevLett.125.236402}%
  \BibitemOpen
  \bibfield  {author} {\bibinfo {author} {\bibfnamefont {D.}~\bibnamefont
  {Rossini}}, \bibinfo {author} {\bibfnamefont {G.~M.}\ \bibnamefont
  {Andolina}}, \bibinfo {author} {\bibfnamefont {D.}~\bibnamefont {Rosa}},
  \bibinfo {author} {\bibfnamefont {M.}~\bibnamefont {Carrega}}, \ and\
  \bibinfo {author} {\bibfnamefont {M.}~\bibnamefont {Polini}},\ }\href
  {\doibase 10.1103/PhysRevLett.125.236402} {\bibfield  {journal} {\bibinfo
  {journal} {Phys. Rev. Lett.}\ }\textbf {\bibinfo {volume} {125}},\ \bibinfo
  {pages} {236402} (\bibinfo {year} {2020})}\BibitemShut {NoStop}%
\bibitem [{\citenamefont {Ito}\ and\ \citenamefont
  {Watanabe}()}]{ito2020collectively}%
  \BibitemOpen
  \bibfield  {author} {\bibinfo {author} {\bibfnamefont {K.}~\bibnamefont
  {Ito}}\ and\ \bibinfo {author} {\bibfnamefont {G.}~\bibnamefont {Watanabe}},\
  }\href@noop {} {}\Eprint {http://arxiv.org/abs/arXiv:2008.07089}
  {arXiv:2008.07089} \BibitemShut {NoStop}%
\bibitem [{\citenamefont {Crescente}\ \emph
  {et~al.}(2020{\natexlab{a}})\citenamefont {Crescente}, \citenamefont
  {Carrega}, \citenamefont {Sassetti},\ and\ \citenamefont
  {Ferraro}}]{PhysRevB.102.245407}%
  \BibitemOpen
  \bibfield  {author} {\bibinfo {author} {\bibfnamefont {A.}~\bibnamefont
  {Crescente}}, \bibinfo {author} {\bibfnamefont {M.}~\bibnamefont {Carrega}},
  \bibinfo {author} {\bibfnamefont {M.}~\bibnamefont {Sassetti}}, \ and\
  \bibinfo {author} {\bibfnamefont {D.}~\bibnamefont {Ferraro}},\ }\href
  {\doibase 10.1103/PhysRevB.102.245407} {\bibfield  {journal} {\bibinfo
  {journal} {Phys. Rev. B}\ }\textbf {\bibinfo {volume} {102}},\ \bibinfo
  {pages} {245407} (\bibinfo {year} {2020}{\natexlab{a}})}\BibitemShut
  {NoStop}%
\bibitem [{\citenamefont {Pirmoradian}\ and\ \citenamefont
  {M\o{}lmer}(2019)}]{PhysRevA.100.043833}%
  \BibitemOpen
  \bibfield  {author} {\bibinfo {author} {\bibfnamefont {F.}~\bibnamefont
  {Pirmoradian}}\ and\ \bibinfo {author} {\bibfnamefont {K.}~\bibnamefont
  {M\o{}lmer}},\ }\href {\doibase 10.1103/PhysRevA.100.043833} {\bibfield
  {journal} {\bibinfo  {journal} {Phys. Rev. A}\ }\textbf {\bibinfo {volume}
  {100}},\ \bibinfo {pages} {043833} (\bibinfo {year} {2019})}\BibitemShut
  {NoStop}%
\bibitem [{\citenamefont {Bai}\ and\ \citenamefont
  {An}(2020)}]{PhysRevA.102.060201}%
  \BibitemOpen
  \bibfield  {author} {\bibinfo {author} {\bibfnamefont {S.-Y.}\ \bibnamefont
  {Bai}}\ and\ \bibinfo {author} {\bibfnamefont {J.-H.}\ \bibnamefont {An}},\
  }\href {\doibase 10.1103/PhysRevA.102.060201} {\bibfield  {journal} {\bibinfo
   {journal} {Phys. Rev. A}\ }\textbf {\bibinfo {volume} {102}},\ \bibinfo
  {pages} {060201(R)} (\bibinfo {year} {2020})}\BibitemShut {NoStop}%
\bibitem [{\citenamefont {Wang}(2021)}]{wang2021fluctuations}%
  \BibitemOpen
  \bibfield  {author} {\bibinfo {author} {\bibfnamefont {S.-Y.}\ \bibnamefont
  {Wang}},\ }\href {\doibase 10.3390/e23111455} {\bibfield  {journal} {\bibinfo
   {journal} {Entropy}\ }\textbf {\bibinfo {volume} {23}},\ \bibinfo {pages}
  {1455} (\bibinfo {year} {2021})}\BibitemShut {NoStop}%
\bibitem [{\citenamefont {Mohan}\ and\ \citenamefont
  {Pati}(2021)}]{PhysRevA.104.042209}%
  \BibitemOpen
  \bibfield  {author} {\bibinfo {author} {\bibfnamefont {B.}~\bibnamefont
  {Mohan}}\ and\ \bibinfo {author} {\bibfnamefont {A.~K.}\ \bibnamefont
  {Pati}},\ }\href {\doibase 10.1103/PhysRevA.104.042209} {\bibfield  {journal}
  {\bibinfo  {journal} {Phys. Rev. A}\ }\textbf {\bibinfo {volume} {104}},\
  \bibinfo {pages} {042209} (\bibinfo {year} {2021})}\BibitemShut {NoStop}%
\bibitem [{\citenamefont {Li}\ \emph {et~al.}(2018)\citenamefont {Li},
  \citenamefont {Lu}, \citenamefont {Chen},\ and\ \citenamefont
  {Amine}}]{Li2018}%
  \BibitemOpen
  \bibfield  {author} {\bibinfo {author} {\bibfnamefont {M.}~\bibnamefont
  {Li}}, \bibinfo {author} {\bibfnamefont {J.}~\bibnamefont {Lu}}, \bibinfo
  {author} {\bibfnamefont {Z.}~\bibnamefont {Chen}}, \ and\ \bibinfo {author}
  {\bibfnamefont {K.}~\bibnamefont {Amine}},\ }\href {\doibase
  10.1002/adma.201800561} {\bibfield  {journal} {\bibinfo  {journal} {Adv.
  Mater.}\ }\textbf {\bibinfo {volume} {30}},\ \bibinfo {pages} {1800561}
  (\bibinfo {year} {2018})}\BibitemShut {NoStop}%
\bibitem [{\citenamefont {Campaioli}\ \emph {et~al.}(2018)\citenamefont
  {Campaioli}, \citenamefont {Pollock},\ and\ \citenamefont
  {Vinjanampathy}}]{campaioli2018quantum}%
  \BibitemOpen
  \bibfield  {author} {\bibinfo {author} {\bibfnamefont {F.}~\bibnamefont
  {Campaioli}}, \bibinfo {author} {\bibfnamefont {F.~A.}\ \bibnamefont
  {Pollock}}, \ and\ \bibinfo {author} {\bibfnamefont {S.}~\bibnamefont
  {Vinjanampathy}},\ }in\ \href@noop {} {\emph {\bibinfo {booktitle}
  {Thermodynamics in the Quantum Regime}}}\ (\bibinfo  {publisher} {Springer},\
  \bibinfo {year} {2018})\ pp.\ \bibinfo {pages} {207--225}\BibitemShut
  {NoStop}%
\bibitem [{\citenamefont {Andolina}\ \emph
  {et~al.}(2019{\natexlab{a}})\citenamefont {Andolina}, \citenamefont {Keck},
  \citenamefont {Mari}, \citenamefont {Giovannetti},\ and\ \citenamefont
  {Polini}}]{PhysRevB.99.205437}%
  \BibitemOpen
  \bibfield  {author} {\bibinfo {author} {\bibfnamefont {G.~M.}\ \bibnamefont
  {Andolina}}, \bibinfo {author} {\bibfnamefont {M.}~\bibnamefont {Keck}},
  \bibinfo {author} {\bibfnamefont {A.}~\bibnamefont {Mari}}, \bibinfo {author}
  {\bibfnamefont {V.}~\bibnamefont {Giovannetti}}, \ and\ \bibinfo {author}
  {\bibfnamefont {M.}~\bibnamefont {Polini}},\ }\href {\doibase
  10.1103/PhysRevB.99.205437} {\bibfield  {journal} {\bibinfo  {journal} {Phys.
  Rev. B}\ }\textbf {\bibinfo {volume} {99}},\ \bibinfo {pages} {205437}
  (\bibinfo {year} {2019}{\natexlab{a}})}\BibitemShut {NoStop}%
\bibitem [{\citenamefont {Cruz}\ \emph {et~al.}(2022)\citenamefont {Cruz},
  \citenamefont {Anka}, \citenamefont {Reis}, \citenamefont {Bachelard},\ and\
  \citenamefont {Santos}}]{Cruz2022}%
  \BibitemOpen
  \bibfield  {author} {\bibinfo {author} {\bibfnamefont {C.}~\bibnamefont
  {Cruz}}, \bibinfo {author} {\bibfnamefont {M.~F.}\ \bibnamefont {Anka}},
  \bibinfo {author} {\bibfnamefont {M.~S.}\ \bibnamefont {Reis}}, \bibinfo
  {author} {\bibfnamefont {R.}~\bibnamefont {Bachelard}}, \ and\ \bibinfo
  {author} {\bibfnamefont {A.~C.}\ \bibnamefont {Santos}},\ }\href {\doibase
  10.1088/2058-9565/ac57f3} {\bibfield  {journal} {\bibinfo  {journal} {Quantum
  Sci. Technol.}\ }\textbf {\bibinfo {volume} {7}},\ \bibinfo {pages} {025020}
  (\bibinfo {year} {2022})}\BibitemShut {NoStop}%
\bibitem [{\citenamefont {Rosa}\ \emph {et~al.}(2020)\citenamefont {Rosa},
  \citenamefont {Rossini}, \citenamefont {Andolina}, \citenamefont {Polini},\
  and\ \citenamefont {Carrega}}]{Rosa2020}%
  \BibitemOpen
  \bibfield  {author} {\bibinfo {author} {\bibfnamefont {D.}~\bibnamefont
  {Rosa}}, \bibinfo {author} {\bibfnamefont {D.}~\bibnamefont {Rossini}},
  \bibinfo {author} {\bibfnamefont {G.~M.}\ \bibnamefont {Andolina}}, \bibinfo
  {author} {\bibfnamefont {M.}~\bibnamefont {Polini}}, \ and\ \bibinfo {author}
  {\bibfnamefont {M.}~\bibnamefont {Carrega}},\ }\href {\doibase
  10.1007/JHEP11(2020)067} {\bibfield  {journal} {\bibinfo  {journal} {J. High
  Energ. Phys.}\ }\textbf {\bibinfo {volume} {2020}},\ \bibinfo {pages} {67}
  (\bibinfo {year} {2020})}\BibitemShut {NoStop}%
\bibitem [{\citenamefont {Moraes}\ \emph {et~al.}(2021)\citenamefont {Moraes},
  \citenamefont {Saguia}, \citenamefont {Santos},\ and\ \citenamefont
  {Sarandy}}]{Moraes2021}%
  \BibitemOpen
  \bibfield  {author} {\bibinfo {author} {\bibfnamefont {L.~F.~C.}\
  \bibnamefont {Moraes}}, \bibinfo {author} {\bibfnamefont {A.}~\bibnamefont
  {Saguia}}, \bibinfo {author} {\bibfnamefont {A.~C.}\ \bibnamefont {Santos}},
  \ and\ \bibinfo {author} {\bibfnamefont {M.~S.}\ \bibnamefont {Sarandy}},\
  }\href {\doibase 10.1209/0295-5075/ac1363} {\bibfield  {journal} {\bibinfo
  {journal} {Europhys. Lett.}\ }\textbf {\bibinfo {volume} {136}},\ \bibinfo
  {pages} {23001} (\bibinfo {year} {2021})}\BibitemShut {NoStop}%
\bibitem [{\citenamefont {Dou}\ \emph {et~al.}(2020)\citenamefont {Dou},
  \citenamefont {Wang},\ and\ \citenamefont {Sun}}]{Dou2020}%
  \BibitemOpen
  \bibfield  {author} {\bibinfo {author} {\bibfnamefont {F.-Q.}\ \bibnamefont
  {Dou}}, \bibinfo {author} {\bibfnamefont {Y.-J.}\ \bibnamefont {Wang}}, \
  and\ \bibinfo {author} {\bibfnamefont {J.-A.}\ \bibnamefont {Sun}},\ }\href
  {\doibase 10.1209/0295-5075/131/43001} {\bibfield  {journal} {\bibinfo
  {journal} {Europhys. Lett.}\ }\textbf {\bibinfo {volume} {131}},\ \bibinfo
  {pages} {43001} (\bibinfo {year} {2020})}\BibitemShut {NoStop}%
\bibitem [{\citenamefont {Dou}\ \emph {et~al.}(2021)\citenamefont {Dou},
  \citenamefont {Wang},\ and\ \citenamefont {Sun}}]{Dou2021}%
  \BibitemOpen
  \bibfield  {author} {\bibinfo {author} {\bibfnamefont {F.-Q.}\ \bibnamefont
  {Dou}}, \bibinfo {author} {\bibfnamefont {Y.-J.}\ \bibnamefont {Wang}}, \
  and\ \bibinfo {author} {\bibfnamefont {J.-A.}\ \bibnamefont {Sun}},\ }\href
  {\doibase 10.1007/s11467-021-1130-5} {\bibfield  {journal} {\bibinfo
  {journal} {Front. Phys.}\ }\textbf {\bibinfo {volume} {17}},\ \bibinfo
  {pages} {31503} (\bibinfo {year} {2021})}\BibitemShut {NoStop}%
\bibitem [{\citenamefont {Dou}\ \emph {et~al.}()\citenamefont {Dou},
  \citenamefont {Wang},\ and\ \citenamefont {Sun}}]{dou2022charging}%
  \BibitemOpen
  \bibfield  {author} {\bibinfo {author} {\bibfnamefont {F.-Q.}\ \bibnamefont
  {Dou}}, \bibinfo {author} {\bibfnamefont {Y.-J.}\ \bibnamefont {Wang}}, \
  and\ \bibinfo {author} {\bibfnamefont {J.-A.}\ \bibnamefont {Sun}},\
  }\href@noop {} {}\Eprint {http://arxiv.org/abs/arXiv:2208.04831}
  {arXiv:2208.04831} \BibitemShut {NoStop}%
\bibitem [{\citenamefont {Santos}\ \emph {et~al.}(2020)\citenamefont {Santos},
  \citenamefont {Saguia},\ and\ \citenamefont {Sarandy}}]{PhysRevE.101.062114}%
  \BibitemOpen
  \bibfield  {author} {\bibinfo {author} {\bibfnamefont {A.~C.}\ \bibnamefont
  {Santos}}, \bibinfo {author} {\bibfnamefont {A.}~\bibnamefont {Saguia}}, \
  and\ \bibinfo {author} {\bibfnamefont {M.~S.}\ \bibnamefont {Sarandy}},\
  }\href {\doibase 10.1103/PhysRevE.101.062114} {\bibfield  {journal} {\bibinfo
   {journal} {Phys. Rev. E}\ }\textbf {\bibinfo {volume} {101}},\ \bibinfo
  {pages} {062114} (\bibinfo {year} {2020})}\BibitemShut {NoStop}%
\bibitem [{\citenamefont {Qi}\ and\ \citenamefont
  {Jing}(2021)}]{PhysRevA.104.032606}%
  \BibitemOpen
  \bibfield  {author} {\bibinfo {author} {\bibfnamefont {S.-F.}\ \bibnamefont
  {Qi}}\ and\ \bibinfo {author} {\bibfnamefont {J.}~\bibnamefont {Jing}},\
  }\href {\doibase 10.1103/PhysRevA.104.032606} {\bibfield  {journal} {\bibinfo
   {journal} {Phys. Rev. A}\ }\textbf {\bibinfo {volume} {104}},\ \bibinfo
  {pages} {032606} (\bibinfo {year} {2021})}\BibitemShut {NoStop}%
\bibitem [{\citenamefont {Dou}\ \emph {et~al.}(2022{\natexlab{a}})\citenamefont
  {Dou}, \citenamefont {Zhou},\ and\ \citenamefont
  {Sun}}]{PhysRevA.106.032212}%
  \BibitemOpen
  \bibfield  {author} {\bibinfo {author} {\bibfnamefont {F.-Q.}\ \bibnamefont
  {Dou}}, \bibinfo {author} {\bibfnamefont {H.}~\bibnamefont {Zhou}}, \ and\
  \bibinfo {author} {\bibfnamefont {J.-A.}\ \bibnamefont {Sun}},\ }\href
  {\doibase 10.1103/PhysRevA.106.032212} {\bibfield  {journal} {\bibinfo
  {journal} {Phys. Rev. A}\ }\textbf {\bibinfo {volume} {106}},\ \bibinfo
  {pages} {032212} (\bibinfo {year} {2022}{\natexlab{a}})}\BibitemShut
  {NoStop}%
\bibitem [{\citenamefont {Zhao}\ \emph {et~al.}(2022)\citenamefont {Zhao},
  \citenamefont {Dou},\ and\ \citenamefont {Zhao}}]{PhysRevResearch.4.013172}%
  \BibitemOpen
  \bibfield  {author} {\bibinfo {author} {\bibfnamefont {F.}~\bibnamefont
  {Zhao}}, \bibinfo {author} {\bibfnamefont {F.-Q.}\ \bibnamefont {Dou}}, \
  and\ \bibinfo {author} {\bibfnamefont {Q.}~\bibnamefont {Zhao}},\ }\href
  {\doibase 10.1103/PhysRevResearch.4.013172} {\bibfield  {journal} {\bibinfo
  {journal} {Phys. Rev. Research}\ }\textbf {\bibinfo {volume} {4}},\ \bibinfo
  {pages} {013172} (\bibinfo {year} {2022})}\BibitemShut {NoStop}%
\bibitem [{\citenamefont {Le}\ \emph {et~al.}(2018)\citenamefont {Le},
  \citenamefont {Levinsen}, \citenamefont {Modi}, \citenamefont {Parish},\ and\
  \citenamefont {Pollock}}]{PhysRevA.97.022106}%
  \BibitemOpen
  \bibfield  {author} {\bibinfo {author} {\bibfnamefont {T.~P.}\ \bibnamefont
  {Le}}, \bibinfo {author} {\bibfnamefont {J.}~\bibnamefont {Levinsen}},
  \bibinfo {author} {\bibfnamefont {K.}~\bibnamefont {Modi}}, \bibinfo {author}
  {\bibfnamefont {M.~M.}\ \bibnamefont {Parish}}, \ and\ \bibinfo {author}
  {\bibfnamefont {F.~A.}\ \bibnamefont {Pollock}},\ }\href {\doibase
  10.1103/PhysRevA.97.022106} {\bibfield  {journal} {\bibinfo  {journal} {Phys.
  Rev. A}\ }\textbf {\bibinfo {volume} {97}},\ \bibinfo {pages} {022106}
  (\bibinfo {year} {2018})}\BibitemShut {NoStop}%
\bibitem [{\citenamefont {Andolina}\ \emph {et~al.}(2018)\citenamefont
  {Andolina}, \citenamefont {Farina}, \citenamefont {Mari}, \citenamefont
  {Pellegrini}, \citenamefont {Giovannetti},\ and\ \citenamefont
  {Polini}}]{PhysRevB.98.205423}%
  \BibitemOpen
  \bibfield  {author} {\bibinfo {author} {\bibfnamefont {G.~M.}\ \bibnamefont
  {Andolina}}, \bibinfo {author} {\bibfnamefont {D.}~\bibnamefont {Farina}},
  \bibinfo {author} {\bibfnamefont {A.}~\bibnamefont {Mari}}, \bibinfo {author}
  {\bibfnamefont {V.}~\bibnamefont {Pellegrini}}, \bibinfo {author}
  {\bibfnamefont {V.}~\bibnamefont {Giovannetti}}, \ and\ \bibinfo {author}
  {\bibfnamefont {M.}~\bibnamefont {Polini}},\ }\href {\doibase
  10.1103/PhysRevB.98.205423} {\bibfield  {journal} {\bibinfo  {journal} {Phys.
  Rev. B}\ }\textbf {\bibinfo {volume} {98}},\ \bibinfo {pages} {205423}
  (\bibinfo {year} {2018})}\BibitemShut {NoStop}%
\bibitem [{\citenamefont {Rossini}\ \emph {et~al.}(2019)\citenamefont
  {Rossini}, \citenamefont {Andolina},\ and\ \citenamefont
  {Polini}}]{PhysRevB.100.115142}%
  \BibitemOpen
  \bibfield  {author} {\bibinfo {author} {\bibfnamefont {D.}~\bibnamefont
  {Rossini}}, \bibinfo {author} {\bibfnamefont {G.~M.}\ \bibnamefont
  {Andolina}}, \ and\ \bibinfo {author} {\bibfnamefont {M.}~\bibnamefont
  {Polini}},\ }\href {\doibase 10.1103/PhysRevB.100.115142} {\bibfield
  {journal} {\bibinfo  {journal} {Phys. Rev. B}\ }\textbf {\bibinfo {volume}
  {100}},\ \bibinfo {pages} {115142} (\bibinfo {year} {2019})}\BibitemShut
  {NoStop}%
\bibitem [{\citenamefont {Peng}\ \emph {et~al.}(2021)\citenamefont {Peng},
  \citenamefont {He}, \citenamefont {Chesi}, \citenamefont {Lin},\ and\
  \citenamefont {Guan}}]{PhysRevA.103.052220}%
  \BibitemOpen
  \bibfield  {author} {\bibinfo {author} {\bibfnamefont {L.}~\bibnamefont
  {Peng}}, \bibinfo {author} {\bibfnamefont {W.-B.}\ \bibnamefont {He}},
  \bibinfo {author} {\bibfnamefont {S.}~\bibnamefont {Chesi}}, \bibinfo
  {author} {\bibfnamefont {H.-Q.}\ \bibnamefont {Lin}}, \ and\ \bibinfo
  {author} {\bibfnamefont {X.-W.}\ \bibnamefont {Guan}},\ }\href {\doibase
  10.1103/PhysRevA.103.052220} {\bibfield  {journal} {\bibinfo  {journal}
  {Phys. Rev. A}\ }\textbf {\bibinfo {volume} {103}},\ \bibinfo {pages}
  {052220} (\bibinfo {year} {2021})}\BibitemShut {NoStop}%
\bibitem [{\citenamefont {Santos}\ \emph {et~al.}(2019)\citenamefont {Santos},
  \citenamefont {Cakmak}, \citenamefont {Campbell},\ and\ \citenamefont
  {Zinner}}]{PhysRevE.100.032107}%
  \BibitemOpen
  \bibfield  {author} {\bibinfo {author} {\bibfnamefont {A.~C.}\ \bibnamefont
  {Santos}}, \bibinfo {author} {\bibfnamefont {B.}~\bibnamefont {Cakmak}},
  \bibinfo {author} {\bibfnamefont {S.}~\bibnamefont {Campbell}}, \ and\
  \bibinfo {author} {\bibfnamefont {N.~T.}\ \bibnamefont {Zinner}},\ }\href
  {\doibase 10.1103/PhysRevE.100.032107} {\bibfield  {journal} {\bibinfo
  {journal} {Phys. Rev. E}\ }\textbf {\bibinfo {volume} {100}},\ \bibinfo
  {pages} {032107} (\bibinfo {year} {2019})}\BibitemShut {NoStop}%
\bibitem [{\citenamefont {Alicki}(2019)}]{Alicki2019}%
  \BibitemOpen
  \bibfield  {author} {\bibinfo {author} {\bibfnamefont {R.}~\bibnamefont
  {Alicki}},\ }\href {\doibase 10.1063/1.5096772} {\bibfield  {journal}
  {\bibinfo  {journal} {J. Chem. Phys.}\ }\textbf {\bibinfo {volume} {150}},\
  \bibinfo {pages} {214110} (\bibinfo {year} {2019})}\BibitemShut {NoStop}%
\bibitem [{\citenamefont {Caravelli}\ \emph {et~al.}(2020)\citenamefont
  {Caravelli}, \citenamefont {Coulter-De~Wit}, \citenamefont
  {Garc\'{\i}a-Pintos},\ and\ \citenamefont
  {Hamma}}]{PhysRevResearch.2.023095}%
  \BibitemOpen
  \bibfield  {author} {\bibinfo {author} {\bibfnamefont {F.}~\bibnamefont
  {Caravelli}}, \bibinfo {author} {\bibfnamefont {G.}~\bibnamefont
  {Coulter-De~Wit}}, \bibinfo {author} {\bibfnamefont {L.~P.}\ \bibnamefont
  {Garc\'{\i}a-Pintos}}, \ and\ \bibinfo {author} {\bibfnamefont
  {A.}~\bibnamefont {Hamma}},\ }\href {\doibase
  10.1103/PhysRevResearch.2.023095} {\bibfield  {journal} {\bibinfo  {journal}
  {Phys. Rev. Research}\ }\textbf {\bibinfo {volume} {2}},\ \bibinfo {pages}
  {023095} (\bibinfo {year} {2020})}\BibitemShut {NoStop}%
\bibitem [{\citenamefont {Chen}\ \emph {et~al.}(2020)\citenamefont {Chen},
  \citenamefont {Zhan}, \citenamefont {Shao}, \citenamefont {Zhang},
  \citenamefont {Zhang},\ and\ \citenamefont {Wang}}]{Chen2020}%
  \BibitemOpen
  \bibfield  {author} {\bibinfo {author} {\bibfnamefont {J.}~\bibnamefont
  {Chen}}, \bibinfo {author} {\bibfnamefont {L.}~\bibnamefont {Zhan}}, \bibinfo
  {author} {\bibfnamefont {L.}~\bibnamefont {Shao}}, \bibinfo {author}
  {\bibfnamefont {X.}~\bibnamefont {Zhang}}, \bibinfo {author} {\bibfnamefont
  {Y.}~\bibnamefont {Zhang}}, \ and\ \bibinfo {author} {\bibfnamefont
  {X.}~\bibnamefont {Wang}},\ }\href {\doibase 10.1002/andp.201900487}
  {\bibfield  {journal} {\bibinfo  {journal} {Ann. Phys.}\ }\textbf {\bibinfo
  {volume} {532}},\ \bibinfo {pages} {1900487} (\bibinfo {year}
  {2020})}\BibitemShut {NoStop}%
\bibitem [{\citenamefont {Konar}\ \emph {et~al.}(2022)\citenamefont {Konar},
  \citenamefont {Lakkaraju}, \citenamefont {Ghosh},\ and\ \citenamefont
  {Sen(De)}}]{PhysRevA.106.022618}%
  \BibitemOpen
  \bibfield  {author} {\bibinfo {author} {\bibfnamefont {T.~K.}\ \bibnamefont
  {Konar}}, \bibinfo {author} {\bibfnamefont {L.~G.~C.}\ \bibnamefont
  {Lakkaraju}}, \bibinfo {author} {\bibfnamefont {S.}~\bibnamefont {Ghosh}}, \
  and\ \bibinfo {author} {\bibfnamefont {A.}~\bibnamefont {Sen(De)}},\ }\href
  {\doibase 10.1103/PhysRevA.106.022618} {\bibfield  {journal} {\bibinfo
  {journal} {Phys. Rev. A}\ }\textbf {\bibinfo {volume} {106}},\ \bibinfo
  {pages} {022618} (\bibinfo {year} {2022})}\BibitemShut {NoStop}%
\bibitem [{\citenamefont {Landi}(2021)}]{landi2021battery}%
  \BibitemOpen
  \bibfield  {author} {\bibinfo {author} {\bibfnamefont {G.~T.}\ \bibnamefont
  {Landi}},\ }\href {\doibase 10.3390/e23121627} {\bibfield  {journal}
  {\bibinfo  {journal} {Entropy}\ }\textbf {\bibinfo {volume} {23}},\ \bibinfo
  {pages} {1627} (\bibinfo {year} {2021})}\BibitemShut {NoStop}%
\bibitem [{\citenamefont {Shaghaghi}\ \emph {et~al.}(2022)\citenamefont
  {Shaghaghi}, \citenamefont {Singh}, \citenamefont {Benenti},\ and\
  \citenamefont {Rosa}}]{Shaghaghi2022}%
  \BibitemOpen
  \bibfield  {author} {\bibinfo {author} {\bibfnamefont {V.}~\bibnamefont
  {Shaghaghi}}, \bibinfo {author} {\bibfnamefont {V.}~\bibnamefont {Singh}},
  \bibinfo {author} {\bibfnamefont {G.}~\bibnamefont {Benenti}}, \ and\
  \bibinfo {author} {\bibfnamefont {D.}~\bibnamefont {Rosa}},\ }\href {\doibase
  10.1088/2058-9565/ac8829} {\bibfield  {journal} {\bibinfo  {journal} {Quantum
  Sci. Technol.}\ }\textbf {\bibinfo {volume} {7}},\ \bibinfo {pages} {04LT01}
  (\bibinfo {year} {2022})}\BibitemShut {NoStop}%
\bibitem [{\citenamefont {Hovhannisyan}\ \emph {et~al.}(2013)\citenamefont
  {Hovhannisyan}, \citenamefont {Perarnau-Llobet}, \citenamefont {Huber},\ and\
  \citenamefont {Ac\'{\i}n}}]{PhysRevLett.111.240401}%
  \BibitemOpen
  \bibfield  {author} {\bibinfo {author} {\bibfnamefont {K.~V.}\ \bibnamefont
  {Hovhannisyan}}, \bibinfo {author} {\bibfnamefont {M.}~\bibnamefont
  {Perarnau-Llobet}}, \bibinfo {author} {\bibfnamefont {M.}~\bibnamefont
  {Huber}}, \ and\ \bibinfo {author} {\bibfnamefont {A.}~\bibnamefont
  {Ac\'{\i}n}},\ }\href {\doibase 10.1103/PhysRevLett.111.240401} {\bibfield
  {journal} {\bibinfo  {journal} {Phys. Rev. Lett.}\ }\textbf {\bibinfo
  {volume} {111}},\ \bibinfo {pages} {240401} (\bibinfo {year}
  {2013})}\BibitemShut {NoStop}%
\bibitem [{\citenamefont {Campaioli}\ \emph {et~al.}(2017)\citenamefont
  {Campaioli}, \citenamefont {Pollock}, \citenamefont {Binder}, \citenamefont
  {C\'eleri}, \citenamefont {Goold}, \citenamefont {Vinjanampathy},\ and\
  \citenamefont {Modi}}]{PhysRevLett.118.150601}%
  \BibitemOpen
  \bibfield  {author} {\bibinfo {author} {\bibfnamefont {F.}~\bibnamefont
  {Campaioli}}, \bibinfo {author} {\bibfnamefont {F.~A.}\ \bibnamefont
  {Pollock}}, \bibinfo {author} {\bibfnamefont {F.~C.}\ \bibnamefont {Binder}},
  \bibinfo {author} {\bibfnamefont {L.}~\bibnamefont {C\'eleri}}, \bibinfo
  {author} {\bibfnamefont {J.}~\bibnamefont {Goold}}, \bibinfo {author}
  {\bibfnamefont {S.}~\bibnamefont {Vinjanampathy}}, \ and\ \bibinfo {author}
  {\bibfnamefont {K.}~\bibnamefont {Modi}},\ }\href {\doibase
  10.1103/PhysRevLett.118.150601} {\bibfield  {journal} {\bibinfo  {journal}
  {Phys. Rev. Lett.}\ }\textbf {\bibinfo {volume} {118}},\ \bibinfo {pages}
  {150601} (\bibinfo {year} {2017})}\BibitemShut {NoStop}%
\bibitem [{\citenamefont {Kamin}\ \emph
  {et~al.}(2020{\natexlab{a}})\citenamefont {Kamin}, \citenamefont {Tabesh},
  \citenamefont {Salimi},\ and\ \citenamefont {Santos}}]{PhysRevE.102.052109}%
  \BibitemOpen
  \bibfield  {author} {\bibinfo {author} {\bibfnamefont {F.~H.}\ \bibnamefont
  {Kamin}}, \bibinfo {author} {\bibfnamefont {F.~T.}\ \bibnamefont {Tabesh}},
  \bibinfo {author} {\bibfnamefont {S.}~\bibnamefont {Salimi}}, \ and\ \bibinfo
  {author} {\bibfnamefont {A.~C.}\ \bibnamefont {Santos}},\ }\href {\doibase
  10.1103/PhysRevE.102.052109} {\bibfield  {journal} {\bibinfo  {journal}
  {Phys. Rev. E}\ }\textbf {\bibinfo {volume} {102}},\ \bibinfo {pages}
  {052109} (\bibinfo {year} {2020}{\natexlab{a}})}\BibitemShut {NoStop}%
\bibitem [{\citenamefont {Sen}\ and\ \citenamefont
  {Sen}(2021)}]{PhysRevA.104.L030402}%
  \BibitemOpen
  \bibfield  {author} {\bibinfo {author} {\bibfnamefont {K.}~\bibnamefont
  {Sen}}\ and\ \bibinfo {author} {\bibfnamefont {U.}~\bibnamefont {Sen}},\
  }\href {\doibase 10.1103/PhysRevA.104.L030402} {\bibfield  {journal}
  {\bibinfo  {journal} {Phys. Rev. A}\ }\textbf {\bibinfo {volume} {104}},\
  \bibinfo {pages} {L030402} (\bibinfo {year} {2021})}\BibitemShut {NoStop}%
\bibitem [{\citenamefont {Liu}\ \emph {et~al.}(2021)\citenamefont {Liu},
  \citenamefont {Shi}, \citenamefont {Shi}, \citenamefont {Wang},\ and\
  \citenamefont {Yang}}]{PhysRevB.104.245418}%
  \BibitemOpen
  \bibfield  {author} {\bibinfo {author} {\bibfnamefont {J.-X.}\ \bibnamefont
  {Liu}}, \bibinfo {author} {\bibfnamefont {H.-L.}\ \bibnamefont {Shi}},
  \bibinfo {author} {\bibfnamefont {Y.-H.}\ \bibnamefont {Shi}}, \bibinfo
  {author} {\bibfnamefont {X.-H.}\ \bibnamefont {Wang}}, \ and\ \bibinfo
  {author} {\bibfnamefont {W.-L.}\ \bibnamefont {Yang}},\ }\href {\doibase
  10.1103/PhysRevB.104.245418} {\bibfield  {journal} {\bibinfo  {journal}
  {Phys. Rev. B}\ }\textbf {\bibinfo {volume} {104}},\ \bibinfo {pages}
  {245418} (\bibinfo {year} {2021})}\BibitemShut {NoStop}%
\bibitem [{\citenamefont {Imai}\ \emph {et~al.}()\citenamefont {Imai},
  \citenamefont {G{\"u}hne},\ and\ \citenamefont {Nimmrichter}}]{imai2022work}%
  \BibitemOpen
  \bibfield  {author} {\bibinfo {author} {\bibfnamefont {S.}~\bibnamefont
  {Imai}}, \bibinfo {author} {\bibfnamefont {O.}~\bibnamefont {G{\"u}hne}}, \
  and\ \bibinfo {author} {\bibfnamefont {S.}~\bibnamefont {Nimmrichter}},\
  }\href@noop {} {}\Eprint {http://arxiv.org/abs/arXiv:2205.08447}
  {arXiv:2205.08447} \BibitemShut {NoStop}%
\bibitem [{\citenamefont {Gyhm}\ \emph {et~al.}(2022)\citenamefont {Gyhm},
  \citenamefont {\ifmmode~\check{S}\else \v{S}\fi{}afr\'anek},\ and\
  \citenamefont {Rosa}}]{PhysRevLett.128.140501}%
  \BibitemOpen
  \bibfield  {author} {\bibinfo {author} {\bibfnamefont {J.-Y.}\ \bibnamefont
  {Gyhm}}, \bibinfo {author} {\bibfnamefont {D.}~\bibnamefont
  {\ifmmode~\check{S}\else \v{S}\fi{}afr\'anek}}, \ and\ \bibinfo {author}
  {\bibfnamefont {D.}~\bibnamefont {Rosa}},\ }\href {\doibase
  10.1103/PhysRevLett.128.140501} {\bibfield  {journal} {\bibinfo  {journal}
  {Phys. Rev. Lett.}\ }\textbf {\bibinfo {volume} {128}},\ \bibinfo {pages}
  {140501} (\bibinfo {year} {2022})}\BibitemShut {NoStop}%
\bibitem [{\citenamefont {Andolina}\ \emph
  {et~al.}(2019{\natexlab{b}})\citenamefont {Andolina}, \citenamefont {Keck},
  \citenamefont {Mari}, \citenamefont {Campisi}, \citenamefont {Giovannetti},\
  and\ \citenamefont {Polini}}]{PhysRevLett.122.047702}%
  \BibitemOpen
  \bibfield  {author} {\bibinfo {author} {\bibfnamefont {G.~M.}\ \bibnamefont
  {Andolina}}, \bibinfo {author} {\bibfnamefont {M.}~\bibnamefont {Keck}},
  \bibinfo {author} {\bibfnamefont {A.}~\bibnamefont {Mari}}, \bibinfo {author}
  {\bibfnamefont {M.}~\bibnamefont {Campisi}}, \bibinfo {author} {\bibfnamefont
  {V.}~\bibnamefont {Giovannetti}}, \ and\ \bibinfo {author} {\bibfnamefont
  {M.}~\bibnamefont {Polini}},\ }\href {\doibase
  10.1103/PhysRevLett.122.047702} {\bibfield  {journal} {\bibinfo  {journal}
  {Phys. Rev. Lett.}\ }\textbf {\bibinfo {volume} {122}},\ \bibinfo {pages}
  {047702} (\bibinfo {year} {2019}{\natexlab{b}})}\BibitemShut {NoStop}%
\bibitem [{\citenamefont {Gumberidze}\ \emph {et~al.}(2019)\citenamefont
  {Gumberidze}, \citenamefont {Kol{\'a}{\v{r}}},\ and\ \citenamefont
  {Filip}}]{Gumberidze2019}%
  \BibitemOpen
  \bibfield  {author} {\bibinfo {author} {\bibfnamefont {M.}~\bibnamefont
  {Gumberidze}}, \bibinfo {author} {\bibfnamefont {M.}~\bibnamefont
  {Kol{\'a}{\v{r}}}}, \ and\ \bibinfo {author} {\bibfnamefont {R.}~\bibnamefont
  {Filip}},\ }\href {\doibase 10.1038/s41598-019-56158-8} {\bibfield  {journal}
  {\bibinfo  {journal} {Sci. Rep.}\ }\textbf {\bibinfo {volume} {9}},\ \bibinfo
  {pages} {19628} (\bibinfo {year} {2019})}\BibitemShut {NoStop}%
\bibitem [{\citenamefont {Centrone}\ \emph {et~al.}()\citenamefont {Centrone},
  \citenamefont {Mancino},\ and\ \citenamefont
  {Paternostro}}]{centrone2021charging}%
  \BibitemOpen
  \bibfield  {author} {\bibinfo {author} {\bibfnamefont {F.}~\bibnamefont
  {Centrone}}, \bibinfo {author} {\bibfnamefont {L.}~\bibnamefont {Mancino}}, \
  and\ \bibinfo {author} {\bibfnamefont {M.}~\bibnamefont {Paternostro}},\
  }\href@noop {} {}\Eprint {http://arxiv.org/abs/arXiv:2106.07899}
  {arXiv:2106.07899} \BibitemShut {NoStop}%
\bibitem [{\citenamefont {Zhang}\ and\ \citenamefont
  {blaauboer}()}]{zhang2018enhanced}%
  \BibitemOpen
  \bibfield  {author} {\bibinfo {author} {\bibfnamefont {X.}~\bibnamefont
  {Zhang}}\ and\ \bibinfo {author} {\bibfnamefont {M.}~\bibnamefont
  {blaauboer}},\ }\href@noop {} {}\Eprint
  {http://arxiv.org/abs/arXiv:1812.10139} {arXiv:1812.10139} \BibitemShut
  {NoStop}%
\bibitem [{\citenamefont {Zhang}\ \emph {et~al.}(2019)\citenamefont {Zhang},
  \citenamefont {Yang}, \citenamefont {Fu},\ and\ \citenamefont
  {Wang}}]{PhysRevE.99.052106}%
  \BibitemOpen
  \bibfield  {author} {\bibinfo {author} {\bibfnamefont {Y.-Y.}\ \bibnamefont
  {Zhang}}, \bibinfo {author} {\bibfnamefont {T.-R.}\ \bibnamefont {Yang}},
  \bibinfo {author} {\bibfnamefont {L.}~\bibnamefont {Fu}}, \ and\ \bibinfo
  {author} {\bibfnamefont {X.}~\bibnamefont {Wang}},\ }\href {\doibase
  10.1103/PhysRevE.99.052106} {\bibfield  {journal} {\bibinfo  {journal} {Phys.
  Rev. E}\ }\textbf {\bibinfo {volume} {99}},\ \bibinfo {pages} {052106}
  (\bibinfo {year} {2019})}\BibitemShut {NoStop}%
\bibitem [{\citenamefont {Ghosh}\ \emph {et~al.}(2020)\citenamefont {Ghosh},
  \citenamefont {Chanda},\ and\ \citenamefont {Sen(De)}}]{PhysRevA.101.032115}%
  \BibitemOpen
  \bibfield  {author} {\bibinfo {author} {\bibfnamefont {S.}~\bibnamefont
  {Ghosh}}, \bibinfo {author} {\bibfnamefont {T.}~\bibnamefont {Chanda}}, \
  and\ \bibinfo {author} {\bibfnamefont {A.}~\bibnamefont {Sen(De)}},\ }\href
  {\doibase 10.1103/PhysRevA.101.032115} {\bibfield  {journal} {\bibinfo
  {journal} {Phys. Rev. A}\ }\textbf {\bibinfo {volume} {101}},\ \bibinfo
  {pages} {032115} (\bibinfo {year} {2020})}\BibitemShut {NoStop}%
\bibitem [{\citenamefont {Huangfu}\ and\ \citenamefont
  {Jing}(2021)}]{PhysRevE.104.024129}%
  \BibitemOpen
  \bibfield  {author} {\bibinfo {author} {\bibfnamefont {Y.}~\bibnamefont
  {Huangfu}}\ and\ \bibinfo {author} {\bibfnamefont {J.}~\bibnamefont {Jing}},\
  }\href {\doibase 10.1103/PhysRevE.104.024129} {\bibfield  {journal} {\bibinfo
   {journal} {Phys. Rev. E}\ }\textbf {\bibinfo {volume} {104}},\ \bibinfo
  {pages} {024129} (\bibinfo {year} {2021})}\BibitemShut {NoStop}%
\bibitem [{\citenamefont {Salvia}\ \emph {et~al.}()\citenamefont {Salvia},
  \citenamefont {Perarnau-Llobet}, \citenamefont {Haack}, \citenamefont
  {Brunner},\ and\ \citenamefont {Nimmrichter}}]{salvia2022quantum}%
  \BibitemOpen
  \bibfield  {author} {\bibinfo {author} {\bibfnamefont {R.}~\bibnamefont
  {Salvia}}, \bibinfo {author} {\bibfnamefont {M.}~\bibnamefont
  {Perarnau-Llobet}}, \bibinfo {author} {\bibfnamefont {G.}~\bibnamefont
  {Haack}}, \bibinfo {author} {\bibfnamefont {N.}~\bibnamefont {Brunner}}, \
  and\ \bibinfo {author} {\bibfnamefont {S.}~\bibnamefont {Nimmrichter}},\
  }\href@noop {} {}\Eprint {http://arxiv.org/abs/arXiv:2205.00026}
  {arXiv:2205.00026} \BibitemShut {NoStop}%
\bibitem [{\citenamefont {Dou}\ \emph {et~al.}(2022{\natexlab{b}})\citenamefont
  {Dou}, \citenamefont {Lu}, \citenamefont {Wang},\ and\ \citenamefont
  {Sun}}]{PhysRevB.105.115405}%
  \BibitemOpen
  \bibfield  {author} {\bibinfo {author} {\bibfnamefont {F.-Q.}\ \bibnamefont
  {Dou}}, \bibinfo {author} {\bibfnamefont {Y.-Q.}\ \bibnamefont {Lu}},
  \bibinfo {author} {\bibfnamefont {Y.-J.}\ \bibnamefont {Wang}}, \ and\
  \bibinfo {author} {\bibfnamefont {J.-A.}\ \bibnamefont {Sun}},\ }\href
  {\doibase 10.1103/PhysRevB.105.115405} {\bibfield  {journal} {\bibinfo
  {journal} {Phys. Rev. B}\ }\textbf {\bibinfo {volume} {105}},\ \bibinfo
  {pages} {115405} (\bibinfo {year} {2022}{\natexlab{b}})}\BibitemShut
  {NoStop}%
\bibitem [{\citenamefont {Kim}\ \emph {et~al.}(2022)\citenamefont {Kim},
  \citenamefont {Murugan}, \citenamefont {Olle},\ and\ \citenamefont
  {Rosa}}]{PhysRevA.105.L010201}%
  \BibitemOpen
  \bibfield  {author} {\bibinfo {author} {\bibfnamefont {J.}~\bibnamefont
  {Kim}}, \bibinfo {author} {\bibfnamefont {J.}~\bibnamefont {Murugan}},
  \bibinfo {author} {\bibfnamefont {J.}~\bibnamefont {Olle}}, \ and\ \bibinfo
  {author} {\bibfnamefont {D.}~\bibnamefont {Rosa}},\ }\href {\doibase
  10.1103/PhysRevA.105.L010201} {\bibfield  {journal} {\bibinfo  {journal}
  {Phys. Rev. A}\ }\textbf {\bibinfo {volume} {105}},\ \bibinfo {pages}
  {L010201} (\bibinfo {year} {2022})}\BibitemShut {NoStop}%
\bibitem [{\citenamefont {Ghosh}\ and\ \citenamefont
  {Sen(De)}(2022)}]{PhysRevA.105.022628}%
  \BibitemOpen
  \bibfield  {author} {\bibinfo {author} {\bibfnamefont {S.}~\bibnamefont
  {Ghosh}}\ and\ \bibinfo {author} {\bibfnamefont {A.}~\bibnamefont
  {Sen(De)}},\ }\href {\doibase 10.1103/PhysRevA.105.022628} {\bibfield
  {journal} {\bibinfo  {journal} {Phys. Rev. A}\ }\textbf {\bibinfo {volume}
  {105}},\ \bibinfo {pages} {022628} (\bibinfo {year} {2022})}\BibitemShut
  {NoStop}%
\bibitem [{\citenamefont {Barra de~la Guarda}(2022)}]{barra2022efficiency}%
  \BibitemOpen
  \bibfield  {author} {\bibinfo {author} {\bibfnamefont {F.}~\bibnamefont
  {Barra de~la Guarda}},\ }\href {\doibase 10.3390/e24060820} {\bibfield
  {journal} {\bibinfo  {journal} {Entropy}\ }\textbf {\bibinfo {volume} {24}},\
  \bibinfo {pages} {820} (\bibinfo {year} {2022})}\BibitemShut {NoStop}%
\bibitem [{\citenamefont {Liu}\ \emph {et~al.}(2019)\citenamefont {Liu},
  \citenamefont {Segal},\ and\ \citenamefont {Hanna}}]{Liu2019}%
  \BibitemOpen
  \bibfield  {author} {\bibinfo {author} {\bibfnamefont {J.}~\bibnamefont
  {Liu}}, \bibinfo {author} {\bibfnamefont {D.}~\bibnamefont {Segal}}, \ and\
  \bibinfo {author} {\bibfnamefont {G.}~\bibnamefont {Hanna}},\ }\href
  {\doibase 10.1021/acs.jpcc.9b06373} {\bibfield  {journal} {\bibinfo
  {journal} {J. Phys. Chem. C}\ }\textbf {\bibinfo {volume} {123}},\ \bibinfo
  {pages} {18303} (\bibinfo {year} {2019})}\BibitemShut {NoStop}%
\bibitem [{\citenamefont {Carrega}\ \emph {et~al.}(2020)\citenamefont
  {Carrega}, \citenamefont {Crescente}, \citenamefont {Ferraro},\ and\
  \citenamefont {Sassetti}}]{Carrega2020}%
  \BibitemOpen
  \bibfield  {author} {\bibinfo {author} {\bibfnamefont {M.}~\bibnamefont
  {Carrega}}, \bibinfo {author} {\bibfnamefont {A.}~\bibnamefont {Crescente}},
  \bibinfo {author} {\bibfnamefont {D.}~\bibnamefont {Ferraro}}, \ and\
  \bibinfo {author} {\bibfnamefont {M.}~\bibnamefont {Sassetti}},\ }\href
  {\doibase 10.1088/1367-2630/abaa01} {\bibfield  {journal} {\bibinfo
  {journal} {New J. Phys.}\ }\textbf {\bibinfo {volume} {22}},\ \bibinfo
  {pages} {083085} (\bibinfo {year} {2020})}\BibitemShut {NoStop}%
\bibitem [{\citenamefont {Xu}\ \emph {et~al.}(2021)\citenamefont {Xu},
  \citenamefont {Zhu}, \citenamefont {Zhang},\ and\ \citenamefont
  {Liu}}]{PhysRevE.104.064143}%
  \BibitemOpen
  \bibfield  {author} {\bibinfo {author} {\bibfnamefont {K.}~\bibnamefont
  {Xu}}, \bibinfo {author} {\bibfnamefont {H.-J.}\ \bibnamefont {Zhu}},
  \bibinfo {author} {\bibfnamefont {G.-F.}\ \bibnamefont {Zhang}}, \ and\
  \bibinfo {author} {\bibfnamefont {W.-M.}\ \bibnamefont {Liu}},\ }\href
  {\doibase 10.1103/PhysRevE.104.064143} {\bibfield  {journal} {\bibinfo
  {journal} {Phys. Rev. E}\ }\textbf {\bibinfo {volume} {104}},\ \bibinfo
  {pages} {064143} (\bibinfo {year} {2021})}\BibitemShut {NoStop}%
\bibitem [{\citenamefont {Ghosh}\ \emph {et~al.}(2021)\citenamefont {Ghosh},
  \citenamefont {Chanda}, \citenamefont {Mal},\ and\ \citenamefont
  {Sen(De)}}]{PhysRevA.104.032207}%
  \BibitemOpen
  \bibfield  {author} {\bibinfo {author} {\bibfnamefont {S.}~\bibnamefont
  {Ghosh}}, \bibinfo {author} {\bibfnamefont {T.}~\bibnamefont {Chanda}},
  \bibinfo {author} {\bibfnamefont {S.}~\bibnamefont {Mal}}, \ and\ \bibinfo
  {author} {\bibfnamefont {A.}~\bibnamefont {Sen(De)}},\ }\href {\doibase
  10.1103/PhysRevA.104.032207} {\bibfield  {journal} {\bibinfo  {journal}
  {Phys. Rev. A}\ }\textbf {\bibinfo {volume} {104}},\ \bibinfo {pages}
  {032207} (\bibinfo {year} {2021})}\BibitemShut {NoStop}%
\bibitem [{\citenamefont {Carrasco}\ \emph {et~al.}(2022)\citenamefont
  {Carrasco}, \citenamefont {Maze}, \citenamefont {Hermann-Avigliano},\ and\
  \citenamefont {Barra}}]{PhysRevE.105.064119}%
  \BibitemOpen
  \bibfield  {author} {\bibinfo {author} {\bibfnamefont {J.}~\bibnamefont
  {Carrasco}}, \bibinfo {author} {\bibfnamefont {J.~R.}\ \bibnamefont {Maze}},
  \bibinfo {author} {\bibfnamefont {C.}~\bibnamefont {Hermann-Avigliano}}, \
  and\ \bibinfo {author} {\bibfnamefont {F.}~\bibnamefont {Barra}},\ }\href
  {\doibase 10.1103/PhysRevE.105.064119} {\bibfield  {journal} {\bibinfo
  {journal} {Phys. Rev. E}\ }\textbf {\bibinfo {volume} {105}},\ \bibinfo
  {pages} {064119} (\bibinfo {year} {2022})}\BibitemShut {NoStop}%
\bibitem [{\citenamefont {Kamin}\ \emph
  {et~al.}(2020{\natexlab{b}})\citenamefont {Kamin}, \citenamefont {Tabesh},
  \citenamefont {Salimi}, \citenamefont {Kheirandish},\ and\ \citenamefont
  {Santos}}]{Kamin2020}%
  \BibitemOpen
  \bibfield  {author} {\bibinfo {author} {\bibfnamefont {F.~H.}\ \bibnamefont
  {Kamin}}, \bibinfo {author} {\bibfnamefont {F.~T.}\ \bibnamefont {Tabesh}},
  \bibinfo {author} {\bibfnamefont {S.}~\bibnamefont {Salimi}}, \bibinfo
  {author} {\bibfnamefont {F.}~\bibnamefont {Kheirandish}}, \ and\ \bibinfo
  {author} {\bibfnamefont {A.~C.}\ \bibnamefont {Santos}},\ }\href {\doibase
  10.1088/1367-2630/ab9ee2} {\bibfield  {journal} {\bibinfo  {journal} {New J.
  Phys.}\ }\textbf {\bibinfo {volume} {22}},\ \bibinfo {pages} {083007}
  (\bibinfo {year} {2020}{\natexlab{b}})}\BibitemShut {NoStop}%
\bibitem [{\citenamefont {Tabesh}\ \emph {et~al.}(2020)\citenamefont {Tabesh},
  \citenamefont {Kamin},\ and\ \citenamefont {Salimi}}]{PhysRevA.102.052223}%
  \BibitemOpen
  \bibfield  {author} {\bibinfo {author} {\bibfnamefont {F.~T.}\ \bibnamefont
  {Tabesh}}, \bibinfo {author} {\bibfnamefont {F.~H.}\ \bibnamefont {Kamin}}, \
  and\ \bibinfo {author} {\bibfnamefont {S.}~\bibnamefont {Salimi}},\ }\href
  {\doibase 10.1103/PhysRevA.102.052223} {\bibfield  {journal} {\bibinfo
  {journal} {Phys. Rev. A}\ }\textbf {\bibinfo {volume} {102}},\ \bibinfo
  {pages} {052223} (\bibinfo {year} {2020})}\BibitemShut {NoStop}%
\bibitem [{\citenamefont {Yao}\ and\ \citenamefont
  {Shao}(2021)}]{PhysRevE.104.044116}%
  \BibitemOpen
  \bibfield  {author} {\bibinfo {author} {\bibfnamefont {Y.}~\bibnamefont
  {Yao}}\ and\ \bibinfo {author} {\bibfnamefont {X.~Q.}\ \bibnamefont {Shao}},\
  }\href {\doibase 10.1103/PhysRevE.104.044116} {\bibfield  {journal} {\bibinfo
   {journal} {Phys. Rev. E}\ }\textbf {\bibinfo {volume} {104}},\ \bibinfo
  {pages} {044116} (\bibinfo {year} {2021})}\BibitemShut {NoStop}%
\bibitem [{\citenamefont {Quach}\ and\ \citenamefont
  {Munro}(2020)}]{PhysRevApplied.14.024092}%
  \BibitemOpen
  \bibfield  {author} {\bibinfo {author} {\bibfnamefont {J.~Q.}\ \bibnamefont
  {Quach}}\ and\ \bibinfo {author} {\bibfnamefont {W.~J.}\ \bibnamefont
  {Munro}},\ }\href {\doibase 10.1103/PhysRevApplied.14.024092} {\bibfield
  {journal} {\bibinfo  {journal} {Phys. Rev. Applied}\ }\textbf {\bibinfo
  {volume} {14}},\ \bibinfo {pages} {024092} (\bibinfo {year}
  {2020})}\BibitemShut {NoStop}%
\bibitem [{\citenamefont {Zhao}\ \emph {et~al.}(2021)\citenamefont {Zhao},
  \citenamefont {Dou},\ and\ \citenamefont {Zhao}}]{PhysRevA.103.033715}%
  \BibitemOpen
  \bibfield  {author} {\bibinfo {author} {\bibfnamefont {F.}~\bibnamefont
  {Zhao}}, \bibinfo {author} {\bibfnamefont {F.-Q.}\ \bibnamefont {Dou}}, \
  and\ \bibinfo {author} {\bibfnamefont {Q.}~\bibnamefont {Zhao}},\ }\href
  {\doibase 10.1103/PhysRevA.103.033715} {\bibfield  {journal} {\bibinfo
  {journal} {Phys. Rev. A}\ }\textbf {\bibinfo {volume} {103}},\ \bibinfo
  {pages} {033715} (\bibinfo {year} {2021})}\BibitemShut {NoStop}%
\bibitem [{\citenamefont {Lu}\ \emph {et~al.}(2021)\citenamefont {Lu},
  \citenamefont {Chen}, \citenamefont {Kuang},\ and\ \citenamefont
  {Wang}}]{PhysRevA.104.043706}%
  \BibitemOpen
  \bibfield  {author} {\bibinfo {author} {\bibfnamefont {W.}~\bibnamefont
  {Lu}}, \bibinfo {author} {\bibfnamefont {J.}~\bibnamefont {Chen}}, \bibinfo
  {author} {\bibfnamefont {L.-M.}\ \bibnamefont {Kuang}}, \ and\ \bibinfo
  {author} {\bibfnamefont {X.}~\bibnamefont {Wang}},\ }\href {\doibase
  10.1103/PhysRevA.104.043706} {\bibfield  {journal} {\bibinfo  {journal}
  {Phys. Rev. A}\ }\textbf {\bibinfo {volume} {104}},\ \bibinfo {pages}
  {043706} (\bibinfo {year} {2021})}\BibitemShut {NoStop}%
\bibitem [{\citenamefont {Farina}\ \emph {et~al.}(2019)\citenamefont {Farina},
  \citenamefont {Andolina}, \citenamefont {Mari}, \citenamefont {Polini},\ and\
  \citenamefont {Giovannetti}}]{PhysRevB.99.035421}%
  \BibitemOpen
  \bibfield  {author} {\bibinfo {author} {\bibfnamefont {D.}~\bibnamefont
  {Farina}}, \bibinfo {author} {\bibfnamefont {G.~M.}\ \bibnamefont
  {Andolina}}, \bibinfo {author} {\bibfnamefont {A.}~\bibnamefont {Mari}},
  \bibinfo {author} {\bibfnamefont {M.}~\bibnamefont {Polini}}, \ and\ \bibinfo
  {author} {\bibfnamefont {V.}~\bibnamefont {Giovannetti}},\ }\href {\doibase
  10.1103/PhysRevB.99.035421} {\bibfield  {journal} {\bibinfo  {journal} {Phys.
  Rev. B}\ }\textbf {\bibinfo {volume} {99}},\ \bibinfo {pages} {035421}
  (\bibinfo {year} {2019})}\BibitemShut {NoStop}%
\bibitem [{\citenamefont {Santos}(2021)}]{PhysRevE.103.042118}%
  \BibitemOpen
  \bibfield  {author} {\bibinfo {author} {\bibfnamefont {A.~C.}\ \bibnamefont
  {Santos}},\ }\href {\doibase 10.1103/PhysRevE.103.042118} {\bibfield
  {journal} {\bibinfo  {journal} {Phys. Rev. E}\ }\textbf {\bibinfo {volume}
  {103}},\ \bibinfo {pages} {042118} (\bibinfo {year} {2021})}\BibitemShut
  {NoStop}%
\bibitem [{\citenamefont {Arjmandi}\ \emph {et~al.}(2022)\citenamefont
  {Arjmandi}, \citenamefont {Mohammadi},\ and\ \citenamefont
  {Santos}}]{PhysRevE.105.054115}%
  \BibitemOpen
  \bibfield  {author} {\bibinfo {author} {\bibfnamefont {M.~B.}\ \bibnamefont
  {Arjmandi}}, \bibinfo {author} {\bibfnamefont {H.}~\bibnamefont {Mohammadi}},
  \ and\ \bibinfo {author} {\bibfnamefont {A.~C.}\ \bibnamefont {Santos}},\
  }\href {\doibase 10.1103/PhysRevE.105.054115} {\bibfield  {journal} {\bibinfo
   {journal} {Phys. Rev. E}\ }\textbf {\bibinfo {volume} {105}},\ \bibinfo
  {pages} {054115} (\bibinfo {year} {2022})}\BibitemShut {NoStop}%
\bibitem [{\citenamefont {Barra}(2019)}]{PhysRevLett.122.210601}%
  \BibitemOpen
  \bibfield  {author} {\bibinfo {author} {\bibfnamefont {F.}~\bibnamefont
  {Barra}},\ }\href {\doibase 10.1103/PhysRevLett.122.210601} {\bibfield
  {journal} {\bibinfo  {journal} {Phys. Rev. Lett.}\ }\textbf {\bibinfo
  {volume} {122}},\ \bibinfo {pages} {210601} (\bibinfo {year}
  {2019})}\BibitemShut {NoStop}%
\bibitem [{\citenamefont {Delmonte}\ \emph {et~al.}(2021)\citenamefont
  {Delmonte}, \citenamefont {Crescente}, \citenamefont {Carrega}, \citenamefont
  {Ferraro},\ and\ \citenamefont {Sassetti}}]{delmonte2021characterization}%
  \BibitemOpen
  \bibfield  {author} {\bibinfo {author} {\bibfnamefont {A.}~\bibnamefont
  {Delmonte}}, \bibinfo {author} {\bibfnamefont {A.}~\bibnamefont {Crescente}},
  \bibinfo {author} {\bibfnamefont {M.}~\bibnamefont {Carrega}}, \bibinfo
  {author} {\bibfnamefont {D.}~\bibnamefont {Ferraro}}, \ and\ \bibinfo
  {author} {\bibfnamefont {M.}~\bibnamefont {Sassetti}},\ }\href {\doibase
  10.3390/e23050612} {\bibfield  {journal} {\bibinfo  {journal} {Entropy}\
  }\textbf {\bibinfo {volume} {23}},\ \bibinfo {pages} {612} (\bibinfo {year}
  {2021})}\BibitemShut {NoStop}%
\bibitem [{\citenamefont {Lipka-Bartosik}\ \emph {et~al.}(2021)\citenamefont
  {Lipka-Bartosik}, \citenamefont {Mazurek},\ and\ \citenamefont
  {Horodecki}}]{LipkaBartosik2021secondlawof}%
  \BibitemOpen
  \bibfield  {author} {\bibinfo {author} {\bibfnamefont {P.}~\bibnamefont
  {Lipka-Bartosik}}, \bibinfo {author} {\bibfnamefont {P.}~\bibnamefont
  {Mazurek}}, \ and\ \bibinfo {author} {\bibfnamefont {M.}~\bibnamefont
  {Horodecki}},\ }\href {\doibase 10.22331/q-2021-03-10-408} {\bibfield
  {journal} {\bibinfo  {journal} {{Quantum}}\ }\textbf {\bibinfo {volume}
  {5}},\ \bibinfo {pages} {408} (\bibinfo {year} {2021})}\BibitemShut {NoStop}%
\bibitem [{\citenamefont {Lu}\ \emph {et~al.}(2022)\citenamefont {Lu},
  \citenamefont {Shao}, \citenamefont {Zhang}, \citenamefont {Zhang},
  \citenamefont {Chen}, \citenamefont {Tao},\ and\ \citenamefont
  {Wang}}]{PhysRevA.105.023718}%
  \BibitemOpen
  \bibfield  {author} {\bibinfo {author} {\bibfnamefont {W.}~\bibnamefont
  {Lu}}, \bibinfo {author} {\bibfnamefont {L.}~\bibnamefont {Shao}}, \bibinfo
  {author} {\bibfnamefont {X.}~\bibnamefont {Zhang}}, \bibinfo {author}
  {\bibfnamefont {Z.}~\bibnamefont {Zhang}}, \bibinfo {author} {\bibfnamefont
  {J.}~\bibnamefont {Chen}}, \bibinfo {author} {\bibfnamefont {H.}~\bibnamefont
  {Tao}}, \ and\ \bibinfo {author} {\bibfnamefont {X.}~\bibnamefont {Wang}},\
  }\href {\doibase 10.1103/PhysRevA.105.023718} {\bibfield  {journal} {\bibinfo
   {journal} {Phys. Rev. A}\ }\textbf {\bibinfo {volume} {105}},\ \bibinfo
  {pages} {023718} (\bibinfo {year} {2022})}\BibitemShut {NoStop}%
\bibitem [{\citenamefont {Monsel}\ \emph {et~al.}(2020)\citenamefont {Monsel},
  \citenamefont {Fellous-Asiani}, \citenamefont {Huard},\ and\ \citenamefont
  {Auff\`eves}}]{PhysRevLett.124.130601}%
  \BibitemOpen
  \bibfield  {author} {\bibinfo {author} {\bibfnamefont {J.}~\bibnamefont
  {Monsel}}, \bibinfo {author} {\bibfnamefont {M.}~\bibnamefont
  {Fellous-Asiani}}, \bibinfo {author} {\bibfnamefont {B.}~\bibnamefont
  {Huard}}, \ and\ \bibinfo {author} {\bibfnamefont {A.}~\bibnamefont
  {Auff\`eves}},\ }\href {\doibase 10.1103/PhysRevLett.124.130601} {\bibfield
  {journal} {\bibinfo  {journal} {Phys. Rev. Lett.}\ }\textbf {\bibinfo
  {volume} {124}},\ \bibinfo {pages} {130601} (\bibinfo {year}
  {2020})}\BibitemShut {NoStop}%
\bibitem [{\citenamefont {Quach}\ \emph {et~al.}(2022)\citenamefont {Quach},
  \citenamefont {McGhee}, \citenamefont {Ganzer}, \citenamefont {Rouse},
  \citenamefont {Lovett}, \citenamefont {Gauger}, \citenamefont {Keeling},
  \citenamefont {Cerullo}, \citenamefont {Lidzey},\ and\ \citenamefont
  {Virgili}}]{quach2020}%
  \BibitemOpen
  \bibfield  {author} {\bibinfo {author} {\bibfnamefont {J.~Q.}\ \bibnamefont
  {Quach}}, \bibinfo {author} {\bibfnamefont {K.~E.}\ \bibnamefont {McGhee}},
  \bibinfo {author} {\bibfnamefont {L.}~\bibnamefont {Ganzer}}, \bibinfo
  {author} {\bibfnamefont {D.~M.}\ \bibnamefont {Rouse}}, \bibinfo {author}
  {\bibfnamefont {B.~W.}\ \bibnamefont {Lovett}}, \bibinfo {author}
  {\bibfnamefont {E.~M.}\ \bibnamefont {Gauger}}, \bibinfo {author}
  {\bibfnamefont {J.}~\bibnamefont {Keeling}}, \bibinfo {author} {\bibfnamefont
  {G.}~\bibnamefont {Cerullo}}, \bibinfo {author} {\bibfnamefont {D.~G.}\
  \bibnamefont {Lidzey}}, \ and\ \bibinfo {author} {\bibfnamefont
  {T.}~\bibnamefont {Virgili}},\ }\href {\doibase 10.1126/sciadv.abk3160}
  {\bibfield  {journal} {\bibinfo  {journal} {Sci. Adv.}\ }\textbf {\bibinfo
  {volume} {8}},\ \bibinfo {pages} {eabk3160} (\bibinfo {year}
  {2022})}\BibitemShut {NoStop}%
\bibitem [{\citenamefont {Hu}\ \emph {et~al.}(2022)\citenamefont {Hu},
  \citenamefont {Qiu}, \citenamefont {Souza}, \citenamefont {Yuan},
  \citenamefont {Zhou}, \citenamefont {Zhang}, \citenamefont {Chu},
  \citenamefont {Pan}, \citenamefont {Hu}, \citenamefont {Li}, \citenamefont
  {Xu}, \citenamefont {Zhong}, \citenamefont {Liu}, \citenamefont {Yan},
  \citenamefont {Tan}, \citenamefont {Bachelard}, \citenamefont {Villas-Boas},
  \citenamefont {Santos},\ and\ \citenamefont {Yu}}]{Hu2022}%
  \BibitemOpen
  \bibfield  {author} {\bibinfo {author} {\bibfnamefont {C.-K.}\ \bibnamefont
  {Hu}}, \bibinfo {author} {\bibfnamefont {J.}~\bibnamefont {Qiu}}, \bibinfo
  {author} {\bibfnamefont {P.~J.~P.}\ \bibnamefont {Souza}}, \bibinfo {author}
  {\bibfnamefont {J.}~\bibnamefont {Yuan}}, \bibinfo {author} {\bibfnamefont
  {Y.}~\bibnamefont {Zhou}}, \bibinfo {author} {\bibfnamefont {L.}~\bibnamefont
  {Zhang}}, \bibinfo {author} {\bibfnamefont {J.}~\bibnamefont {Chu}}, \bibinfo
  {author} {\bibfnamefont {X.}~\bibnamefont {Pan}}, \bibinfo {author}
  {\bibfnamefont {L.}~\bibnamefont {Hu}}, \bibinfo {author} {\bibfnamefont
  {J.}~\bibnamefont {Li}}, \bibinfo {author} {\bibfnamefont {Y.}~\bibnamefont
  {Xu}}, \bibinfo {author} {\bibfnamefont {Y.}~\bibnamefont {Zhong}}, \bibinfo
  {author} {\bibfnamefont {S.}~\bibnamefont {Liu}}, \bibinfo {author}
  {\bibfnamefont {F.}~\bibnamefont {Yan}}, \bibinfo {author} {\bibfnamefont
  {D.}~\bibnamefont {Tan}}, \bibinfo {author} {\bibfnamefont {R.}~\bibnamefont
  {Bachelard}}, \bibinfo {author} {\bibfnamefont {C.~J.}\ \bibnamefont
  {Villas-Boas}}, \bibinfo {author} {\bibfnamefont {A.~C.}\ \bibnamefont
  {Santos}}, \ and\ \bibinfo {author} {\bibfnamefont {D.}~\bibnamefont {Yu}},\
  }\href {\doibase 10.1088/2058-9565/ac8444} {\bibfield  {journal} {\bibinfo
  {journal} {Quantum Sci. Technol.}\ }\textbf {\bibinfo {volume} {7}},\
  \bibinfo {pages} {045018} (\bibinfo {year} {2022})}\BibitemShut {NoStop}%
\bibitem [{\citenamefont {Joshi}\ and\ \citenamefont
  {Mahesh}(2022)}]{PhysRevA.106.042601}%
  \BibitemOpen
  \bibfield  {author} {\bibinfo {author} {\bibfnamefont {J.}~\bibnamefont
  {Joshi}}\ and\ \bibinfo {author} {\bibfnamefont {T.~S.}\ \bibnamefont
  {Mahesh}},\ }\href {\doibase 10.1103/PhysRevA.106.042601} {\bibfield
  {journal} {\bibinfo  {journal} {Phys. Rev. A}\ }\textbf {\bibinfo {volume}
  {106}},\ \bibinfo {pages} {042601} (\bibinfo {year} {2022})}\BibitemShut
  {NoStop}%
\bibitem [{\citenamefont {Wenniger}\ \emph {et~al.}()\citenamefont {Wenniger},
  \citenamefont {Thomas}, \citenamefont {Maffei}, \citenamefont {Wein},
  \citenamefont {Pont}, \citenamefont {Harouri}, \citenamefont {Lema{\^\i}tre},
  \citenamefont {Sagnes}, \citenamefont {Somaschi}, \citenamefont
  {Auff{\`e}ves} \emph {et~al.}}]{wenniger2022coherence}%
  \BibitemOpen
  \bibfield  {author} {\bibinfo {author} {\bibfnamefont {I.}~\bibnamefont
  {Wenniger}}, \bibinfo {author} {\bibfnamefont {S.}~\bibnamefont {Thomas}},
  \bibinfo {author} {\bibfnamefont {M.}~\bibnamefont {Maffei}}, \bibinfo
  {author} {\bibfnamefont {S.}~\bibnamefont {Wein}}, \bibinfo {author}
  {\bibfnamefont {M.}~\bibnamefont {Pont}}, \bibinfo {author} {\bibfnamefont
  {A.}~\bibnamefont {Harouri}}, \bibinfo {author} {\bibfnamefont
  {A.}~\bibnamefont {Lema{\^\i}tre}}, \bibinfo {author} {\bibfnamefont
  {I.}~\bibnamefont {Sagnes}}, \bibinfo {author} {\bibfnamefont
  {N.}~\bibnamefont {Somaschi}}, \bibinfo {author} {\bibfnamefont
  {A.}~\bibnamefont {Auff{\`e}ves}},  \emph {et~al.},\ }\href@noop {} {}\Eprint
  {http://arxiv.org/abs/arXiv:2202.01109} {arXiv:2202.01109} \BibitemShut
  {NoStop}%
\bibitem [{\citenamefont {Gemme}\ \emph {et~al.}(2022)\citenamefont {Gemme},
  \citenamefont {Grossi}, \citenamefont {Ferraro}, \citenamefont {Vallecorsa},\
  and\ \citenamefont {Sassetti}}]{gemme2022ibm}%
  \BibitemOpen
  \bibfield  {author} {\bibinfo {author} {\bibfnamefont {G.}~\bibnamefont
  {Gemme}}, \bibinfo {author} {\bibfnamefont {M.}~\bibnamefont {Grossi}},
  \bibinfo {author} {\bibfnamefont {D.}~\bibnamefont {Ferraro}}, \bibinfo
  {author} {\bibfnamefont {S.}~\bibnamefont {Vallecorsa}}, \ and\ \bibinfo
  {author} {\bibfnamefont {M.}~\bibnamefont {Sassetti}},\ }\href {\doibase
  10.3390/batteries8050043} {\bibfield  {journal} {\bibinfo  {journal}
  {Batteries}\ }\textbf {\bibinfo {volume} {8}},\ \bibinfo {pages} {43}
  (\bibinfo {year} {2022})}\BibitemShut {NoStop}%
\bibitem [{\citenamefont {Zheng}\ \emph {et~al.}(2022)\citenamefont {Zheng},
  \citenamefont {Ning}, \citenamefont {Yang}, \citenamefont {Xia},\ and\
  \citenamefont {Zheng}}]{Zheng2022}%
  \BibitemOpen
  \bibfield  {author} {\bibinfo {author} {\bibfnamefont {R.-H.}\ \bibnamefont
  {Zheng}}, \bibinfo {author} {\bibfnamefont {W.}~\bibnamefont {Ning}},
  \bibinfo {author} {\bibfnamefont {Z.-B.}\ \bibnamefont {Yang}}, \bibinfo
  {author} {\bibfnamefont {Y.}~\bibnamefont {Xia}}, \ and\ \bibinfo {author}
  {\bibfnamefont {S.-B.}\ \bibnamefont {Zheng}},\ }\href {\doibase
  10.1088/1367-2630/ac788f} {\bibfield  {journal} {\bibinfo  {journal} {New J.
  Phys.}\ }\textbf {\bibinfo {volume} {24}},\ \bibinfo {pages} {063031}
  (\bibinfo {year} {2022})}\BibitemShut {NoStop}%
\bibitem [{\citenamefont {Crescente}\ \emph {et~al.}(2022)\citenamefont
  {Crescente}, \citenamefont {Ferraro}, \citenamefont {Carrega},\ and\
  \citenamefont {Sassetti}}]{PhysRevResearch.4.033216}%
  \BibitemOpen
  \bibfield  {author} {\bibinfo {author} {\bibfnamefont {A.}~\bibnamefont
  {Crescente}}, \bibinfo {author} {\bibfnamefont {D.}~\bibnamefont {Ferraro}},
  \bibinfo {author} {\bibfnamefont {M.}~\bibnamefont {Carrega}}, \ and\
  \bibinfo {author} {\bibfnamefont {M.}~\bibnamefont {Sassetti}},\ }\href
  {\doibase 10.1103/PhysRevResearch.4.033216} {\bibfield  {journal} {\bibinfo
  {journal} {Phys. Rev. Research}\ }\textbf {\bibinfo {volume} {4}},\ \bibinfo
  {pages} {033216} (\bibinfo {year} {2022})}\BibitemShut {NoStop}%
\bibitem [{\citenamefont {You}\ and\ \citenamefont {Nori}(2005)}]{You2005}%
  \BibitemOpen
  \bibfield  {author} {\bibinfo {author} {\bibfnamefont {J.~Q.}\ \bibnamefont
  {You}}\ and\ \bibinfo {author} {\bibfnamefont {F.}~\bibnamefont {Nori}},\
  }\href {\doibase 10.1063/1.2155757} {\bibfield  {journal} {\bibinfo
  {journal} {Phys. Today}\ }\textbf {\bibinfo {volume} {58}},\ \bibinfo {pages}
  {42} (\bibinfo {year} {2005})}\BibitemShut {NoStop}%
\bibitem [{\citenamefont {Devoret}\ and\ \citenamefont
  {Schoelkopf}(2013)}]{H.2013}%
  \BibitemOpen
  \bibfield  {author} {\bibinfo {author} {\bibfnamefont {M.~H.}\ \bibnamefont
  {Devoret}}\ and\ \bibinfo {author} {\bibfnamefont {R.~J.}\ \bibnamefont
  {Schoelkopf}},\ }\href {\doibase 10.1126/science.1231930} {\bibfield
  {journal} {\bibinfo  {journal} {Science}\ }\textbf {\bibinfo {volume}
  {339}},\ \bibinfo {pages} {1169} (\bibinfo {year} {2013})}\BibitemShut
  {NoStop}%
\bibitem [{\citenamefont {Xiang}\ \emph {et~al.}(2013)\citenamefont {Xiang},
  \citenamefont {Ashhab}, \citenamefont {You},\ and\ \citenamefont
  {Nori}}]{RevModPhys.85.623}%
  \BibitemOpen
  \bibfield  {author} {\bibinfo {author} {\bibfnamefont {Z.-L.}\ \bibnamefont
  {Xiang}}, \bibinfo {author} {\bibfnamefont {S.}~\bibnamefont {Ashhab}},
  \bibinfo {author} {\bibfnamefont {J.~Q.}\ \bibnamefont {You}}, \ and\
  \bibinfo {author} {\bibfnamefont {F.}~\bibnamefont {Nori}},\ }\href {\doibase
  10.1103/RevModPhys.85.623} {\bibfield  {journal} {\bibinfo  {journal} {Rev.
  Mod. Phys.}\ }\textbf {\bibinfo {volume} {85}},\ \bibinfo {pages} {623}
  (\bibinfo {year} {2013})}\BibitemShut {NoStop}%
\bibitem [{\citenamefont {Gu}\ \emph {et~al.}(2017)\citenamefont {Gu},
  \citenamefont {Kockum}, \citenamefont {Miranowicz}, \citenamefont {Liu},\
  and\ \citenamefont {Nori}}]{Gu2017}%
  \BibitemOpen
  \bibfield  {author} {\bibinfo {author} {\bibfnamefont {X.}~\bibnamefont
  {Gu}}, \bibinfo {author} {\bibfnamefont {A.~F.}\ \bibnamefont {Kockum}},
  \bibinfo {author} {\bibfnamefont {A.}~\bibnamefont {Miranowicz}}, \bibinfo
  {author} {\bibfnamefont {Y.-X.}\ \bibnamefont {Liu}}, \ and\ \bibinfo
  {author} {\bibfnamefont {F.}~\bibnamefont {Nori}},\ }\href
  {https://www.sciencedirect.com/science/article/pii/S0370157317303290}
  {\bibfield  {journal} {\bibinfo  {journal} {Phys. Rep.}\ }\textbf {\bibinfo
  {volume} {718-719}},\ \bibinfo {pages} {1} (\bibinfo {year}
  {2017})}\BibitemShut {NoStop}%
\bibitem [{\citenamefont {Gao}\ \emph {et~al.}(2021)\citenamefont {Gao},
  \citenamefont {Rol}, \citenamefont {Touzard},\ and\ \citenamefont
  {Wang}}]{PRXQuantum.2.040202}%
  \BibitemOpen
  \bibfield  {author} {\bibinfo {author} {\bibfnamefont {Y.~Y.}\ \bibnamefont
  {Gao}}, \bibinfo {author} {\bibfnamefont {M.~A.}\ \bibnamefont {Rol}},
  \bibinfo {author} {\bibfnamefont {S.}~\bibnamefont {Touzard}}, \ and\
  \bibinfo {author} {\bibfnamefont {C.}~\bibnamefont {Wang}},\ }\href {\doibase
  10.1103/PRXQuantum.2.040202} {\bibfield  {journal} {\bibinfo  {journal} {PRX
  Quantum}\ }\textbf {\bibinfo {volume} {2}},\ \bibinfo {pages} {040202}
  (\bibinfo {year} {2021})}\BibitemShut {NoStop}%
\bibitem [{\citenamefont {Strambini}\ \emph {et~al.}(2020)\citenamefont
  {Strambini}, \citenamefont {Iorio}, \citenamefont {Durante}, \citenamefont
  {Citro}, \citenamefont {Sanz-Fern{\'a}ndez}, \citenamefont {Guarcello},
  \citenamefont {Tokatly}, \citenamefont {Braggio}, \citenamefont {Rocci},
  \citenamefont {Ligato}, \citenamefont {Zannier}, \citenamefont {Sorba},
  \citenamefont {Bergeret},\ and\ \citenamefont {Giazotto}}]{Strambini2020}%
  \BibitemOpen
  \bibfield  {author} {\bibinfo {author} {\bibfnamefont {E.}~\bibnamefont
  {Strambini}}, \bibinfo {author} {\bibfnamefont {A.}~\bibnamefont {Iorio}},
  \bibinfo {author} {\bibfnamefont {O.}~\bibnamefont {Durante}}, \bibinfo
  {author} {\bibfnamefont {R.}~\bibnamefont {Citro}}, \bibinfo {author}
  {\bibfnamefont {C.}~\bibnamefont {Sanz-Fern{\'a}ndez}}, \bibinfo {author}
  {\bibfnamefont {C.}~\bibnamefont {Guarcello}}, \bibinfo {author}
  {\bibfnamefont {I.~V.}\ \bibnamefont {Tokatly}}, \bibinfo {author}
  {\bibfnamefont {A.}~\bibnamefont {Braggio}}, \bibinfo {author} {\bibfnamefont
  {M.}~\bibnamefont {Rocci}}, \bibinfo {author} {\bibfnamefont
  {N.}~\bibnamefont {Ligato}}, \bibinfo {author} {\bibfnamefont
  {V.}~\bibnamefont {Zannier}}, \bibinfo {author} {\bibfnamefont
  {L.}~\bibnamefont {Sorba}}, \bibinfo {author} {\bibfnamefont {F.~S.}\
  \bibnamefont {Bergeret}}, \ and\ \bibinfo {author} {\bibfnamefont
  {F.}~\bibnamefont {Giazotto}},\ }\href {\doibase 10.1038/s41565-020-0712-7}
  {\bibfield  {journal} {\bibinfo  {journal} {Nat. Nanotechnol.}\ }\textbf
  {\bibinfo {volume} {15}},\ \bibinfo {pages} {656} (\bibinfo {year}
  {2020})}\BibitemShut {NoStop}%
\bibitem [{\citenamefont {Crescente}\ \emph
  {et~al.}(2020{\natexlab{b}})\citenamefont {Crescente}, \citenamefont
  {Carrega}, \citenamefont {Sassetti},\ and\ \citenamefont
  {Ferraro}}]{Crescente2020}%
  \BibitemOpen
  \bibfield  {author} {\bibinfo {author} {\bibfnamefont {A.}~\bibnamefont
  {Crescente}}, \bibinfo {author} {\bibfnamefont {M.}~\bibnamefont {Carrega}},
  \bibinfo {author} {\bibfnamefont {M.}~\bibnamefont {Sassetti}}, \ and\
  \bibinfo {author} {\bibfnamefont {D.}~\bibnamefont {Ferraro}},\ }\href
  {\doibase 10.1088/1367-2630/ab91fc} {\bibfield  {journal} {\bibinfo
  {journal} {New J. Phys.}\ }\textbf {\bibinfo {volume} {22}},\ \bibinfo
  {pages} {063057} (\bibinfo {year} {2020}{\natexlab{b}})}\BibitemShut
  {NoStop}%
\bibitem [{\citenamefont {Masuki}\ \emph {et~al.}(2022)\citenamefont {Masuki},
  \citenamefont {Sudo}, \citenamefont {Oshikawa},\ and\ \citenamefont
  {Ashida}}]{PhysRevLett.129.087001}%
  \BibitemOpen
  \bibfield  {author} {\bibinfo {author} {\bibfnamefont {K.}~\bibnamefont
  {Masuki}}, \bibinfo {author} {\bibfnamefont {H.}~\bibnamefont {Sudo}},
  \bibinfo {author} {\bibfnamefont {M.}~\bibnamefont {Oshikawa}}, \ and\
  \bibinfo {author} {\bibfnamefont {Y.}~\bibnamefont {Ashida}},\ }\href
  {\doibase 10.1103/PhysRevLett.129.087001} {\bibfield  {journal} {\bibinfo
  {journal} {Phys. Rev. Lett.}\ }\textbf {\bibinfo {volume} {129}},\ \bibinfo
  {pages} {087001} (\bibinfo {year} {2022})}\BibitemShut {NoStop}%
\bibitem [{\citenamefont {Jaako}\ \emph {et~al.}(2016)\citenamefont {Jaako},
  \citenamefont {Xiang}, \citenamefont {Garcia-Ripoll},\ and\ \citenamefont
  {Rabl}}]{PhysRevA.94.033850}%
  \BibitemOpen
  \bibfield  {author} {\bibinfo {author} {\bibfnamefont {T.}~\bibnamefont
  {Jaako}}, \bibinfo {author} {\bibfnamefont {Z.-L.}\ \bibnamefont {Xiang}},
  \bibinfo {author} {\bibfnamefont {J.~J.}\ \bibnamefont {Garcia-Ripoll}}, \
  and\ \bibinfo {author} {\bibfnamefont {P.}~\bibnamefont {Rabl}},\ }\href
  {\doibase 10.1103/PhysRevA.94.033850} {\bibfield  {journal} {\bibinfo
  {journal} {Phys. Rev. A}\ }\textbf {\bibinfo {volume} {94}},\ \bibinfo
  {pages} {033850} (\bibinfo {year} {2016})}\BibitemShut {NoStop}%
\bibitem [{\citenamefont {Kaur}\ \emph {et~al.}(2021)\citenamefont {Kaur},
  \citenamefont {S\'epulcre}, \citenamefont {Roch}, \citenamefont {Snyman},
  \citenamefont {Florens},\ and\ \citenamefont
  {Bera}}]{PhysRevLett.127.237702}%
  \BibitemOpen
  \bibfield  {author} {\bibinfo {author} {\bibfnamefont {K.}~\bibnamefont
  {Kaur}}, \bibinfo {author} {\bibfnamefont {T.}~\bibnamefont {S\'epulcre}},
  \bibinfo {author} {\bibfnamefont {N.}~\bibnamefont {Roch}}, \bibinfo {author}
  {\bibfnamefont {I.}~\bibnamefont {Snyman}}, \bibinfo {author} {\bibfnamefont
  {S.}~\bibnamefont {Florens}}, \ and\ \bibinfo {author} {\bibfnamefont
  {S.}~\bibnamefont {Bera}},\ }\href {\doibase 10.1103/PhysRevLett.127.237702}
  {\bibfield  {journal} {\bibinfo  {journal} {Phys. Rev. Lett.}\ }\textbf
  {\bibinfo {volume} {127}},\ \bibinfo {pages} {237702} (\bibinfo {year}
  {2021})}\BibitemShut {NoStop}%
\bibitem [{\citenamefont {Blais}\ \emph {et~al.}(2004)\citenamefont {Blais},
  \citenamefont {Huang}, \citenamefont {Wallraff}, \citenamefont {Girvin},\
  and\ \citenamefont {Schoelkopf}}]{PhysRevA.69.062320}%
  \BibitemOpen
  \bibfield  {author} {\bibinfo {author} {\bibfnamefont {A.}~\bibnamefont
  {Blais}}, \bibinfo {author} {\bibfnamefont {R.-S.}\ \bibnamefont {Huang}},
  \bibinfo {author} {\bibfnamefont {A.}~\bibnamefont {Wallraff}}, \bibinfo
  {author} {\bibfnamefont {S.~M.}\ \bibnamefont {Girvin}}, \ and\ \bibinfo
  {author} {\bibfnamefont {R.~J.}\ \bibnamefont {Schoelkopf}},\ }\href
  {\doibase 10.1103/PhysRevA.69.062320} {\bibfield  {journal} {\bibinfo
  {journal} {Phys. Rev. A}\ }\textbf {\bibinfo {volume} {69}},\ \bibinfo
  {pages} {062320} (\bibinfo {year} {2004})}\BibitemShut {NoStop}%
\bibitem [{\citenamefont {Blais}\ \emph {et~al.}(2021)\citenamefont {Blais},
  \citenamefont {Grimsmo}, \citenamefont {Girvin},\ and\ \citenamefont
  {Wallraff}}]{RevModPhys.93.025005}%
  \BibitemOpen
  \bibfield  {author} {\bibinfo {author} {\bibfnamefont {A.}~\bibnamefont
  {Blais}}, \bibinfo {author} {\bibfnamefont {A.~L.}\ \bibnamefont {Grimsmo}},
  \bibinfo {author} {\bibfnamefont {S.~M.}\ \bibnamefont {Girvin}}, \ and\
  \bibinfo {author} {\bibfnamefont {A.}~\bibnamefont {Wallraff}},\ }\href
  {\doibase 10.1103/RevModPhys.93.025005} {\bibfield  {journal} {\bibinfo
  {journal} {Rev. Mod. Phys.}\ }\textbf {\bibinfo {volume} {93}},\ \bibinfo
  {pages} {025005} (\bibinfo {year} {2021})}\BibitemShut {NoStop}%
\bibitem [{\citenamefont {Koch}\ \emph {et~al.}(2007)\citenamefont {Koch},
  \citenamefont {Yu}, \citenamefont {Gambetta}, \citenamefont {Houck},
  \citenamefont {Schuster}, \citenamefont {Majer}, \citenamefont {Blais},
  \citenamefont {Devoret}, \citenamefont {Girvin},\ and\ \citenamefont
  {Schoelkopf}}]{PhysRevA.76.042319}%
  \BibitemOpen
  \bibfield  {author} {\bibinfo {author} {\bibfnamefont {J.}~\bibnamefont
  {Koch}}, \bibinfo {author} {\bibfnamefont {T.~M.}\ \bibnamefont {Yu}},
  \bibinfo {author} {\bibfnamefont {J.}~\bibnamefont {Gambetta}}, \bibinfo
  {author} {\bibfnamefont {A.~A.}\ \bibnamefont {Houck}}, \bibinfo {author}
  {\bibfnamefont {D.~I.}\ \bibnamefont {Schuster}}, \bibinfo {author}
  {\bibfnamefont {J.}~\bibnamefont {Majer}}, \bibinfo {author} {\bibfnamefont
  {A.}~\bibnamefont {Blais}}, \bibinfo {author} {\bibfnamefont {M.~H.}\
  \bibnamefont {Devoret}}, \bibinfo {author} {\bibfnamefont {S.~M.}\
  \bibnamefont {Girvin}}, \ and\ \bibinfo {author} {\bibfnamefont {R.~J.}\
  \bibnamefont {Schoelkopf}},\ }\href {\doibase 10.1103/PhysRevA.76.042319}
  {\bibfield  {journal} {\bibinfo  {journal} {Phys. Rev. A}\ }\textbf {\bibinfo
  {volume} {76}},\ \bibinfo {pages} {042319} (\bibinfo {year}
  {2007})}\BibitemShut {NoStop}%
\bibitem [{\citenamefont {Zhang}\ \emph {et~al.}(2022)\citenamefont {Zhang},
  \citenamefont {Curtis}, \citenamefont {Wang}, \citenamefont {Schoelkopf},\
  and\ \citenamefont {Girvin}}]{PhysRevA.105.022423}%
  \BibitemOpen
  \bibfield  {author} {\bibinfo {author} {\bibfnamefont {Y.}~\bibnamefont
  {Zhang}}, \bibinfo {author} {\bibfnamefont {J.~C.}\ \bibnamefont {Curtis}},
  \bibinfo {author} {\bibfnamefont {C.~S.}\ \bibnamefont {Wang}}, \bibinfo
  {author} {\bibfnamefont {R.~J.}\ \bibnamefont {Schoelkopf}}, \ and\ \bibinfo
  {author} {\bibfnamefont {S.~M.}\ \bibnamefont {Girvin}},\ }\href {\doibase
  10.1103/PhysRevA.105.022423} {\bibfield  {journal} {\bibinfo  {journal}
  {Phys. Rev. A}\ }\textbf {\bibinfo {volume} {105}},\ \bibinfo {pages}
  {022423} (\bibinfo {year} {2022})}\BibitemShut {NoStop}%
\bibitem [{\citenamefont {Gely}\ and\ \citenamefont
  {Steele}(2021)}]{PhysRevA.104.053509}%
  \BibitemOpen
  \bibfield  {author} {\bibinfo {author} {\bibfnamefont {M.~F.}\ \bibnamefont
  {Gely}}\ and\ \bibinfo {author} {\bibfnamefont {G.~A.}\ \bibnamefont
  {Steele}},\ }\href {\doibase 10.1103/PhysRevA.104.053509} {\bibfield
  {journal} {\bibinfo  {journal} {Phys. Rev. A}\ }\textbf {\bibinfo {volume}
  {104}},\ \bibinfo {pages} {053509} (\bibinfo {year} {2021})}\BibitemShut
  {NoStop}%
\bibitem [{\citenamefont {Zhang}\ \emph {et~al.}(2013)\citenamefont {Zhang},
  \citenamefont {Wu}, \citenamefont {Li}, \citenamefont {Dai},\ and\
  \citenamefont {Li}}]{PhysRevA.87.062325}%
  \BibitemOpen
  \bibfield  {author} {\bibinfo {author} {\bibfnamefont {Z.-R.}\ \bibnamefont
  {Zhang}}, \bibinfo {author} {\bibfnamefont {C.-W.}\ \bibnamefont {Wu}},
  \bibinfo {author} {\bibfnamefont {C.-Y.}\ \bibnamefont {Li}}, \bibinfo
  {author} {\bibfnamefont {H.-Y.}\ \bibnamefont {Dai}}, \ and\ \bibinfo
  {author} {\bibfnamefont {C.-Z.}\ \bibnamefont {Li}},\ }\href {\doibase
  10.1103/PhysRevA.87.062325} {\bibfield  {journal} {\bibinfo  {journal} {Phys.
  Rev. A}\ }\textbf {\bibinfo {volume} {87}},\ \bibinfo {pages} {062325}
  (\bibinfo {year} {2013})}\BibitemShut {NoStop}%
\bibitem [{\citenamefont {Schreier}\ \emph {et~al.}(2008)\citenamefont
  {Schreier}, \citenamefont {Houck}, \citenamefont {Koch}, \citenamefont
  {Schuster}, \citenamefont {Johnson}, \citenamefont {Chow}, \citenamefont
  {Gambetta}, \citenamefont {Majer}, \citenamefont {Frunzio}, \citenamefont
  {Devoret}, \citenamefont {Girvin},\ and\ \citenamefont
  {Schoelkopf}}]{PhysRevB.77.180502}%
  \BibitemOpen
  \bibfield  {author} {\bibinfo {author} {\bibfnamefont {J.~A.}\ \bibnamefont
  {Schreier}}, \bibinfo {author} {\bibfnamefont {A.~A.}\ \bibnamefont {Houck}},
  \bibinfo {author} {\bibfnamefont {J.}~\bibnamefont {Koch}}, \bibinfo {author}
  {\bibfnamefont {D.~I.}\ \bibnamefont {Schuster}}, \bibinfo {author}
  {\bibfnamefont {B.~R.}\ \bibnamefont {Johnson}}, \bibinfo {author}
  {\bibfnamefont {J.~M.}\ \bibnamefont {Chow}}, \bibinfo {author}
  {\bibfnamefont {J.~M.}\ \bibnamefont {Gambetta}}, \bibinfo {author}
  {\bibfnamefont {J.}~\bibnamefont {Majer}}, \bibinfo {author} {\bibfnamefont
  {L.}~\bibnamefont {Frunzio}}, \bibinfo {author} {\bibfnamefont {M.~H.}\
  \bibnamefont {Devoret}}, \bibinfo {author} {\bibfnamefont {S.~M.}\
  \bibnamefont {Girvin}}, \ and\ \bibinfo {author} {\bibfnamefont {R.~J.}\
  \bibnamefont {Schoelkopf}},\ }\href {\doibase 10.1103/PhysRevB.77.180502}
  {\bibfield  {journal} {\bibinfo  {journal} {Phys. Rev. B}\ }\textbf {\bibinfo
  {volume} {77}},\ \bibinfo {pages} {180502} (\bibinfo {year}
  {2008})}\BibitemShut {NoStop}%
\bibitem [{\citenamefont {Houck}\ \emph {et~al.}(2009)\citenamefont {Houck},
  \citenamefont {Koch}, \citenamefont {Devoret}, \citenamefont {Girvin},\ and\
  \citenamefont {Schoelkopf}}]{Houck2009}%
  \BibitemOpen
  \bibfield  {author} {\bibinfo {author} {\bibfnamefont {A.~A.}\ \bibnamefont
  {Houck}}, \bibinfo {author} {\bibfnamefont {J.}~\bibnamefont {Koch}},
  \bibinfo {author} {\bibfnamefont {M.~H.}\ \bibnamefont {Devoret}}, \bibinfo
  {author} {\bibfnamefont {S.~M.}\ \bibnamefont {Girvin}}, \ and\ \bibinfo
  {author} {\bibfnamefont {R.~J.}\ \bibnamefont {Schoelkopf}},\ }\href
  {\doibase 10.1007/s11128-009-0100-6} {\bibfield  {journal} {\bibinfo
  {journal} {Quantum Inf. Proc.}\ }\textbf {\bibinfo {volume} {8}},\ \bibinfo
  {pages} {105} (\bibinfo {year} {2009})}\BibitemShut {NoStop}%
\bibitem [{\citenamefont {Forn-D\'{\i}az}\ \emph {et~al.}(2019)\citenamefont
  {Forn-D\'{\i}az}, \citenamefont {Lamata}, \citenamefont {Rico}, \citenamefont
  {Kono},\ and\ \citenamefont {Solano}}]{RevModPhys.91.025005}%
  \BibitemOpen
  \bibfield  {author} {\bibinfo {author} {\bibfnamefont {P.}~\bibnamefont
  {Forn-D\'{\i}az}}, \bibinfo {author} {\bibfnamefont {L.}~\bibnamefont
  {Lamata}}, \bibinfo {author} {\bibfnamefont {E.}~\bibnamefont {Rico}},
  \bibinfo {author} {\bibfnamefont {J.}~\bibnamefont {Kono}}, \ and\ \bibinfo
  {author} {\bibfnamefont {E.}~\bibnamefont {Solano}},\ }\href {\doibase
  10.1103/RevModPhys.91.025005} {\bibfield  {journal} {\bibinfo  {journal}
  {Rev. Mod. Phys.}\ }\textbf {\bibinfo {volume} {91}},\ \bibinfo {pages}
  {025005} (\bibinfo {year} {2019})}\BibitemShut {NoStop}%
\bibitem [{\citenamefont {Krantz}\ \emph {et~al.}(2019)\citenamefont {Krantz},
  \citenamefont {Kjaergaard}, \citenamefont {Yan}, \citenamefont {Orlando},
  \citenamefont {Gustavsson},\ and\ \citenamefont {Oliver}}]{Krantz2019}%
  \BibitemOpen
  \bibfield  {author} {\bibinfo {author} {\bibfnamefont {P.}~\bibnamefont
  {Krantz}}, \bibinfo {author} {\bibfnamefont {M.}~\bibnamefont {Kjaergaard}},
  \bibinfo {author} {\bibfnamefont {F.}~\bibnamefont {Yan}}, \bibinfo {author}
  {\bibfnamefont {T.~P.}\ \bibnamefont {Orlando}}, \bibinfo {author}
  {\bibfnamefont {S.}~\bibnamefont {Gustavsson}}, \ and\ \bibinfo {author}
  {\bibfnamefont {W.~D.}\ \bibnamefont {Oliver}},\ }\href {\doibase
  10.1063/1.5089550} {\bibfield  {journal} {\bibinfo  {journal} {Appl. Phys.
  Rev.}\ }\textbf {\bibinfo {volume} {6}},\ \bibinfo {pages} {021318} (\bibinfo
  {year} {2019})}\BibitemShut {NoStop}%
\bibitem [{\citenamefont {Schuster}\ \emph {et~al.}(2007)\citenamefont
  {Schuster}, \citenamefont {Houck}, \citenamefont {Schreier}, \citenamefont
  {Wallraff}, \citenamefont {Gambetta}, \citenamefont {Blais}, \citenamefont
  {Frunzio}, \citenamefont {Majer}, \citenamefont {Johnson}, \citenamefont
  {Devoret}, \citenamefont {Girvin},\ and\ \citenamefont
  {Schoelkopf}}]{Schuster2007}%
  \BibitemOpen
  \bibfield  {author} {\bibinfo {author} {\bibfnamefont {D.~I.}\ \bibnamefont
  {Schuster}}, \bibinfo {author} {\bibfnamefont {A.~A.}\ \bibnamefont {Houck}},
  \bibinfo {author} {\bibfnamefont {J.~A.}\ \bibnamefont {Schreier}}, \bibinfo
  {author} {\bibfnamefont {A.}~\bibnamefont {Wallraff}}, \bibinfo {author}
  {\bibfnamefont {J.~M.}\ \bibnamefont {Gambetta}}, \bibinfo {author}
  {\bibfnamefont {A.}~\bibnamefont {Blais}}, \bibinfo {author} {\bibfnamefont
  {L.}~\bibnamefont {Frunzio}}, \bibinfo {author} {\bibfnamefont
  {J.}~\bibnamefont {Majer}}, \bibinfo {author} {\bibfnamefont
  {B.}~\bibnamefont {Johnson}}, \bibinfo {author} {\bibfnamefont {M.~H.}\
  \bibnamefont {Devoret}}, \bibinfo {author} {\bibfnamefont {S.~M.}\
  \bibnamefont {Girvin}}, \ and\ \bibinfo {author} {\bibfnamefont {R.~J.}\
  \bibnamefont {Schoelkopf}},\ }\href {\doibase 10.1038/nature05461} {\bibfield
   {journal} {\bibinfo  {journal} {Nature (London)}\ }\textbf {\bibinfo
  {volume} {445}},\ \bibinfo {pages} {515} (\bibinfo {year}
  {2007})}\BibitemShut {NoStop}%
\bibitem [{\citenamefont {Bishop}\ \emph {et~al.}(2009)\citenamefont {Bishop},
  \citenamefont {Chow}, \citenamefont {Koch}, \citenamefont {Houck},
  \citenamefont {Devoret}, \citenamefont {Thuneberg}, \citenamefont {Girvin},\
  and\ \citenamefont {Schoelkopf}}]{Bishop2009}%
  \BibitemOpen
  \bibfield  {author} {\bibinfo {author} {\bibfnamefont {L.~S.}\ \bibnamefont
  {Bishop}}, \bibinfo {author} {\bibfnamefont {J.~M.}\ \bibnamefont {Chow}},
  \bibinfo {author} {\bibfnamefont {J.}~\bibnamefont {Koch}}, \bibinfo {author}
  {\bibfnamefont {A.~A.}\ \bibnamefont {Houck}}, \bibinfo {author}
  {\bibfnamefont {M.~H.}\ \bibnamefont {Devoret}}, \bibinfo {author}
  {\bibfnamefont {E.}~\bibnamefont {Thuneberg}}, \bibinfo {author}
  {\bibfnamefont {S.~M.}\ \bibnamefont {Girvin}}, \ and\ \bibinfo {author}
  {\bibfnamefont {R.~J.}\ \bibnamefont {Schoelkopf}},\ }\href {\doibase
  10.1038/nphys1154} {\bibfield  {journal} {\bibinfo  {journal} {Nat. Phys.}\
  }\textbf {\bibinfo {volume} {5}},\ \bibinfo {pages} {105} (\bibinfo {year}
  {2009})}\BibitemShut {NoStop}%
\bibitem [{\citenamefont {Bosman}\ \emph
  {et~al.}(2017{\natexlab{a}})\citenamefont {Bosman}, \citenamefont {Gely},
  \citenamefont {Singh}, \citenamefont {Bothner}, \citenamefont
  {Castellanos-Gomez},\ and\ \citenamefont {Steele}}]{PhysRevB.95.224515}%
  \BibitemOpen
  \bibfield  {author} {\bibinfo {author} {\bibfnamefont {S.~J.}\ \bibnamefont
  {Bosman}}, \bibinfo {author} {\bibfnamefont {M.~F.}\ \bibnamefont {Gely}},
  \bibinfo {author} {\bibfnamefont {V.}~\bibnamefont {Singh}}, \bibinfo
  {author} {\bibfnamefont {D.}~\bibnamefont {Bothner}}, \bibinfo {author}
  {\bibfnamefont {A.}~\bibnamefont {Castellanos-Gomez}}, \ and\ \bibinfo
  {author} {\bibfnamefont {G.~A.}\ \bibnamefont {Steele}},\ }\href {\doibase
  10.1103/PhysRevB.95.224515} {\bibfield  {journal} {\bibinfo  {journal} {Phys.
  Rev. B}\ }\textbf {\bibinfo {volume} {95}},\ \bibinfo {pages} {224515}
  (\bibinfo {year} {2017}{\natexlab{a}})}\BibitemShut {NoStop}%
\bibitem [{\citenamefont {Bosman}\ \emph
  {et~al.}(2017{\natexlab{b}})\citenamefont {Bosman}, \citenamefont {Gely},
  \citenamefont {Singh}, \citenamefont {Bruno}, \citenamefont {Bothner},\ and\
  \citenamefont {Steele}}]{Bosman2017}%
  \BibitemOpen
  \bibfield  {author} {\bibinfo {author} {\bibfnamefont {S.~J.}\ \bibnamefont
  {Bosman}}, \bibinfo {author} {\bibfnamefont {M.~F.}\ \bibnamefont {Gely}},
  \bibinfo {author} {\bibfnamefont {V.}~\bibnamefont {Singh}}, \bibinfo
  {author} {\bibfnamefont {A.}~\bibnamefont {Bruno}}, \bibinfo {author}
  {\bibfnamefont {D.}~\bibnamefont {Bothner}}, \ and\ \bibinfo {author}
  {\bibfnamefont {G.~A.}\ \bibnamefont {Steele}},\ }\href {\doibase
  10.1038/s41534-017-0046-y} {\bibfield  {journal} {\bibinfo  {journal} {npj
  Quantum Inf.}\ }\textbf {\bibinfo {volume} {3}},\ \bibinfo {pages} {46}
  (\bibinfo {year} {2017}{\natexlab{b}})}\BibitemShut {NoStop}%
\bibitem [{\citenamefont {Abah}\ \emph {et~al.}(2022)\citenamefont {Abah},
  \citenamefont {De~Chiara}, \citenamefont {Paternostro},\ and\ \citenamefont
  {Puebla}}]{PhysRevResearch.4.L022017}%
  \BibitemOpen
  \bibfield  {author} {\bibinfo {author} {\bibfnamefont {O.}~\bibnamefont
  {Abah}}, \bibinfo {author} {\bibfnamefont {G.}~\bibnamefont {De~Chiara}},
  \bibinfo {author} {\bibfnamefont {M.}~\bibnamefont {Paternostro}}, \ and\
  \bibinfo {author} {\bibfnamefont {R.}~\bibnamefont {Puebla}},\ }\href
  {\doibase 10.1103/PhysRevResearch.4.L022017} {\bibfield  {journal} {\bibinfo
  {journal} {Phys. Rev. Research}\ }\textbf {\bibinfo {volume} {4}},\ \bibinfo
  {pages} {L022017} (\bibinfo {year} {2022})}\BibitemShut {NoStop}%
\bibitem [{\citenamefont {Wallraff}\ \emph {et~al.}(2004)\citenamefont
  {Wallraff}, \citenamefont {Schuster}, \citenamefont {Blais}, \citenamefont
  {Frunzio}, \citenamefont {Huang}, \citenamefont {Majer}, \citenamefont
  {Kumar}, \citenamefont {Girvin},\ and\ \citenamefont
  {Schoelkopf}}]{Wallraff2004}%
  \BibitemOpen
  \bibfield  {author} {\bibinfo {author} {\bibfnamefont {A.}~\bibnamefont
  {Wallraff}}, \bibinfo {author} {\bibfnamefont {D.~I.}\ \bibnamefont
  {Schuster}}, \bibinfo {author} {\bibfnamefont {A.}~\bibnamefont {Blais}},
  \bibinfo {author} {\bibfnamefont {L.}~\bibnamefont {Frunzio}}, \bibinfo
  {author} {\bibfnamefont {R.-S.}\ \bibnamefont {Huang}}, \bibinfo {author}
  {\bibfnamefont {J.}~\bibnamefont {Majer}}, \bibinfo {author} {\bibfnamefont
  {S.}~\bibnamefont {Kumar}}, \bibinfo {author} {\bibfnamefont {S.~M.}\
  \bibnamefont {Girvin}}, \ and\ \bibinfo {author} {\bibfnamefont {R.~J.}\
  \bibnamefont {Schoelkopf}},\ }\href {\doibase 10.1038/nature02851} {\bibfield
   {journal} {\bibinfo  {journal} {Nature (London)}\ }\textbf {\bibinfo
  {volume} {431}},\ \bibinfo {pages} {162} (\bibinfo {year}
  {2004})}\BibitemShut {NoStop}%
\bibitem [{\citenamefont {Niemczyk}\ \emph {et~al.}(2010)\citenamefont
  {Niemczyk}, \citenamefont {Deppe}, \citenamefont {Huebl}, \citenamefont
  {Menzel}, \citenamefont {Hocke}, \citenamefont {Schwarz}, \citenamefont
  {Garcia-Ripoll}, \citenamefont {Zueco}, \citenamefont {H{\"u}mmer},
  \citenamefont {Solano}, \citenamefont {Marx},\ and\ \citenamefont
  {Gross}}]{Niemczyk2010}%
  \BibitemOpen
  \bibfield  {author} {\bibinfo {author} {\bibfnamefont {T.}~\bibnamefont
  {Niemczyk}}, \bibinfo {author} {\bibfnamefont {F.}~\bibnamefont {Deppe}},
  \bibinfo {author} {\bibfnamefont {H.}~\bibnamefont {Huebl}}, \bibinfo
  {author} {\bibfnamefont {E.~P.}\ \bibnamefont {Menzel}}, \bibinfo {author}
  {\bibfnamefont {F.}~\bibnamefont {Hocke}}, \bibinfo {author} {\bibfnamefont
  {M.~J.}\ \bibnamefont {Schwarz}}, \bibinfo {author} {\bibfnamefont {J.~J.}\
  \bibnamefont {Garcia-Ripoll}}, \bibinfo {author} {\bibfnamefont
  {D.}~\bibnamefont {Zueco}}, \bibinfo {author} {\bibfnamefont
  {T.}~\bibnamefont {H{\"u}mmer}}, \bibinfo {author} {\bibfnamefont
  {E.}~\bibnamefont {Solano}}, \bibinfo {author} {\bibfnamefont
  {A.}~\bibnamefont {Marx}}, \ and\ \bibinfo {author} {\bibfnamefont
  {R.}~\bibnamefont {Gross}},\ }\href {\doibase 10.1038/nphys1730} {\bibfield
  {journal} {\bibinfo  {journal} {Nat. Phys.}\ }\textbf {\bibinfo {volume}
  {6}},\ \bibinfo {pages} {772} (\bibinfo {year} {2010})}\BibitemShut {NoStop}%
\bibitem [{\citenamefont {Forn-D\'{\i}az}\ \emph {et~al.}(2010)\citenamefont
  {Forn-D\'{\i}az}, \citenamefont {Lisenfeld}, \citenamefont {Marcos},
  \citenamefont {Garc\'{\i}a-Ripoll}, \citenamefont {Solano}, \citenamefont
  {Harmans},\ and\ \citenamefont {Mooij}}]{PhysRevLett.105.237001}%
  \BibitemOpen
  \bibfield  {author} {\bibinfo {author} {\bibfnamefont {P.}~\bibnamefont
  {Forn-D\'{\i}az}}, \bibinfo {author} {\bibfnamefont {J.}~\bibnamefont
  {Lisenfeld}}, \bibinfo {author} {\bibfnamefont {D.}~\bibnamefont {Marcos}},
  \bibinfo {author} {\bibfnamefont {J.~J.}\ \bibnamefont {Garc\'{\i}a-Ripoll}},
  \bibinfo {author} {\bibfnamefont {E.}~\bibnamefont {Solano}}, \bibinfo
  {author} {\bibfnamefont {C.~J. P.~M.}\ \bibnamefont {Harmans}}, \ and\
  \bibinfo {author} {\bibfnamefont {J.~E.}\ \bibnamefont {Mooij}},\ }\href
  {\doibase 10.1103/PhysRevLett.105.237001} {\bibfield  {journal} {\bibinfo
  {journal} {Phys. Rev. Lett.}\ }\textbf {\bibinfo {volume} {105}},\ \bibinfo
  {pages} {237001} (\bibinfo {year} {2010})}\BibitemShut {NoStop}%
\bibitem [{\citenamefont {Baust}\ \emph {et~al.}(2016)\citenamefont {Baust},
  \citenamefont {Hoffmann}, \citenamefont {Haeberlein}, \citenamefont
  {Schwarz}, \citenamefont {Eder}, \citenamefont {Goetz}, \citenamefont
  {Wulschner}, \citenamefont {Xie}, \citenamefont {Zhong}, \citenamefont
  {Quijandr\'{\i}a}, \citenamefont {Zueco}, \citenamefont {Ripoll},
  \citenamefont {Garc\'{\i}a-\'Alvarez}, \citenamefont {Romero}, \citenamefont
  {Solano}, \citenamefont {Fedorov}, \citenamefont {Menzel}, \citenamefont
  {Deppe}, \citenamefont {Marx},\ and\ \citenamefont
  {Gross}}]{PhysRevB.93.214501}%
  \BibitemOpen
  \bibfield  {author} {\bibinfo {author} {\bibfnamefont {A.}~\bibnamefont
  {Baust}}, \bibinfo {author} {\bibfnamefont {E.}~\bibnamefont {Hoffmann}},
  \bibinfo {author} {\bibfnamefont {M.}~\bibnamefont {Haeberlein}}, \bibinfo
  {author} {\bibfnamefont {M.~J.}\ \bibnamefont {Schwarz}}, \bibinfo {author}
  {\bibfnamefont {P.}~\bibnamefont {Eder}}, \bibinfo {author} {\bibfnamefont
  {J.}~\bibnamefont {Goetz}}, \bibinfo {author} {\bibfnamefont
  {F.}~\bibnamefont {Wulschner}}, \bibinfo {author} {\bibfnamefont
  {E.}~\bibnamefont {Xie}}, \bibinfo {author} {\bibfnamefont {L.}~\bibnamefont
  {Zhong}}, \bibinfo {author} {\bibfnamefont {F.}~\bibnamefont
  {Quijandr\'{\i}a}}, \bibinfo {author} {\bibfnamefont {D.}~\bibnamefont
  {Zueco}}, \bibinfo {author} {\bibfnamefont {J.-J.~G.}\ \bibnamefont
  {Ripoll}}, \bibinfo {author} {\bibfnamefont {L.}~\bibnamefont
  {Garc\'{\i}a-\'Alvarez}}, \bibinfo {author} {\bibfnamefont {G.}~\bibnamefont
  {Romero}}, \bibinfo {author} {\bibfnamefont {E.}~\bibnamefont {Solano}},
  \bibinfo {author} {\bibfnamefont {K.~G.}\ \bibnamefont {Fedorov}}, \bibinfo
  {author} {\bibfnamefont {E.~P.}\ \bibnamefont {Menzel}}, \bibinfo {author}
  {\bibfnamefont {F.}~\bibnamefont {Deppe}}, \bibinfo {author} {\bibfnamefont
  {A.}~\bibnamefont {Marx}}, \ and\ \bibinfo {author} {\bibfnamefont
  {R.}~\bibnamefont {Gross}},\ }\href {\doibase 10.1103/PhysRevB.93.214501}
  {\bibfield  {journal} {\bibinfo  {journal} {Phys. Rev. B}\ }\textbf {\bibinfo
  {volume} {93}},\ \bibinfo {pages} {214501} (\bibinfo {year}
  {2016})}\BibitemShut {NoStop}%
\bibitem [{\citenamefont {Yoshihara}\ \emph
  {et~al.}(2017{\natexlab{a}})\citenamefont {Yoshihara}, \citenamefont {Fuse},
  \citenamefont {Ashhab}, \citenamefont {Kakuyanagi}, \citenamefont {Saito},\
  and\ \citenamefont {Semba}}]{PhysRevA.95.053824}%
  \BibitemOpen
  \bibfield  {author} {\bibinfo {author} {\bibfnamefont {F.}~\bibnamefont
  {Yoshihara}}, \bibinfo {author} {\bibfnamefont {T.}~\bibnamefont {Fuse}},
  \bibinfo {author} {\bibfnamefont {S.}~\bibnamefont {Ashhab}}, \bibinfo
  {author} {\bibfnamefont {K.}~\bibnamefont {Kakuyanagi}}, \bibinfo {author}
  {\bibfnamefont {S.}~\bibnamefont {Saito}}, \ and\ \bibinfo {author}
  {\bibfnamefont {K.}~\bibnamefont {Semba}},\ }\href {\doibase
  10.1103/PhysRevA.95.053824} {\bibfield  {journal} {\bibinfo  {journal} {Phys.
  Rev. A}\ }\textbf {\bibinfo {volume} {95}},\ \bibinfo {pages} {053824}
  (\bibinfo {year} {2017}{\natexlab{a}})}\BibitemShut {NoStop}%
\bibitem [{\citenamefont {Yoshihara}\ \emph
  {et~al.}(2017{\natexlab{b}})\citenamefont {Yoshihara}, \citenamefont {Fuse},
  \citenamefont {Ashhab}, \citenamefont {Kakuyanagi}, \citenamefont {Saito},\
  and\ \citenamefont {Semba}}]{Yoshihara2017}%
  \BibitemOpen
  \bibfield  {author} {\bibinfo {author} {\bibfnamefont {F.}~\bibnamefont
  {Yoshihara}}, \bibinfo {author} {\bibfnamefont {T.}~\bibnamefont {Fuse}},
  \bibinfo {author} {\bibfnamefont {S.}~\bibnamefont {Ashhab}}, \bibinfo
  {author} {\bibfnamefont {K.}~\bibnamefont {Kakuyanagi}}, \bibinfo {author}
  {\bibfnamefont {S.}~\bibnamefont {Saito}}, \ and\ \bibinfo {author}
  {\bibfnamefont {K.}~\bibnamefont {Semba}},\ }\href {\doibase
  10.1038/nphys3906} {\bibfield  {journal} {\bibinfo  {journal} {Nat. Phys.}\
  }\textbf {\bibinfo {volume} {13}},\ \bibinfo {pages} {44} (\bibinfo {year}
  {2017}{\natexlab{b}})}\BibitemShut {NoStop}%
\bibitem [{\citenamefont {Yoshihara}\ \emph {et~al.}(2018)\citenamefont
  {Yoshihara}, \citenamefont {Fuse}, \citenamefont {Ao}, \citenamefont
  {Ashhab}, \citenamefont {Kakuyanagi}, \citenamefont {Saito}, \citenamefont
  {Aoki}, \citenamefont {Koshino},\ and\ \citenamefont
  {Semba}}]{PhysRevLett.120.183601}%
  \BibitemOpen
  \bibfield  {author} {\bibinfo {author} {\bibfnamefont {F.}~\bibnamefont
  {Yoshihara}}, \bibinfo {author} {\bibfnamefont {T.}~\bibnamefont {Fuse}},
  \bibinfo {author} {\bibfnamefont {Z.}~\bibnamefont {Ao}}, \bibinfo {author}
  {\bibfnamefont {S.}~\bibnamefont {Ashhab}}, \bibinfo {author} {\bibfnamefont
  {K.}~\bibnamefont {Kakuyanagi}}, \bibinfo {author} {\bibfnamefont
  {S.}~\bibnamefont {Saito}}, \bibinfo {author} {\bibfnamefont
  {T.}~\bibnamefont {Aoki}}, \bibinfo {author} {\bibfnamefont {K.}~\bibnamefont
  {Koshino}}, \ and\ \bibinfo {author} {\bibfnamefont {K.}~\bibnamefont
  {Semba}},\ }\href {\doibase 10.1103/PhysRevLett.120.183601} {\bibfield
  {journal} {\bibinfo  {journal} {Phys. Rev. Lett.}\ }\textbf {\bibinfo
  {volume} {120}},\ \bibinfo {pages} {183601} (\bibinfo {year}
  {2018})}\BibitemShut {NoStop}%
\bibitem [{\citenamefont {Forn-D{\'i}az}\ \emph {et~al.}(2017)\citenamefont
  {Forn-D{\'i}az}, \citenamefont {Garc{\'i}a-Ripoll}, \citenamefont
  {Peropadre}, \citenamefont {Orgiazzi}, \citenamefont {Yurtalan},
  \citenamefont {Belyansky}, \citenamefont {Wilson},\ and\ \citenamefont
  {Lupascu}}]{Forn-Daz2017}%
  \BibitemOpen
  \bibfield  {author} {\bibinfo {author} {\bibfnamefont {P.}~\bibnamefont
  {Forn-D{\'i}az}}, \bibinfo {author} {\bibfnamefont {J.~J.}\ \bibnamefont
  {Garc{\'i}a-Ripoll}}, \bibinfo {author} {\bibfnamefont {B.}~\bibnamefont
  {Peropadre}}, \bibinfo {author} {\bibfnamefont {J.-L.}\ \bibnamefont
  {Orgiazzi}}, \bibinfo {author} {\bibfnamefont {M.~A.}\ \bibnamefont
  {Yurtalan}}, \bibinfo {author} {\bibfnamefont {R.}~\bibnamefont {Belyansky}},
  \bibinfo {author} {\bibfnamefont {C.~M.}\ \bibnamefont {Wilson}}, \ and\
  \bibinfo {author} {\bibfnamefont {A.}~\bibnamefont {Lupascu}},\ }\href
  {\doibase 10.1038/nphys3905} {\bibfield  {journal} {\bibinfo  {journal} {Nat.
  Phys.}\ }\textbf {\bibinfo {volume} {13}},\ \bibinfo {pages} {39} (\bibinfo
  {year} {2017})}\BibitemShut {NoStop}%
\bibitem [{\citenamefont {Fink}\ \emph {et~al.}(2009)\citenamefont {Fink},
  \citenamefont {Bianchetti}, \citenamefont {Baur}, \citenamefont {G\"oppl},
  \citenamefont {Steffen}, \citenamefont {Filipp}, \citenamefont {Leek},
  \citenamefont {Blais},\ and\ \citenamefont
  {Wallraff}}]{PhysRevLett.103.083601}%
  \BibitemOpen
  \bibfield  {author} {\bibinfo {author} {\bibfnamefont {J.~M.}\ \bibnamefont
  {Fink}}, \bibinfo {author} {\bibfnamefont {R.}~\bibnamefont {Bianchetti}},
  \bibinfo {author} {\bibfnamefont {M.}~\bibnamefont {Baur}}, \bibinfo {author}
  {\bibfnamefont {M.}~\bibnamefont {G\"oppl}}, \bibinfo {author} {\bibfnamefont
  {L.}~\bibnamefont {Steffen}}, \bibinfo {author} {\bibfnamefont
  {S.}~\bibnamefont {Filipp}}, \bibinfo {author} {\bibfnamefont {P.~J.}\
  \bibnamefont {Leek}}, \bibinfo {author} {\bibfnamefont {A.}~\bibnamefont
  {Blais}}, \ and\ \bibinfo {author} {\bibfnamefont {A.}~\bibnamefont
  {Wallraff}},\ }\href {\doibase 10.1103/PhysRevLett.103.083601} {\bibfield
  {journal} {\bibinfo  {journal} {Phys. Rev. Lett.}\ }\textbf {\bibinfo
  {volume} {103}},\ \bibinfo {pages} {083601} (\bibinfo {year}
  {2009})}\BibitemShut {NoStop}%
\bibitem [{\citenamefont {Langford}\ \emph {et~al.}(2017)\citenamefont
  {Langford}, \citenamefont {Sagastizabal}, \citenamefont {Kounalakis},
  \citenamefont {Dickel}, \citenamefont {Bruno}, \citenamefont {Luthi},
  \citenamefont {Thoen}, \citenamefont {Endo},\ and\ \citenamefont
  {DiCarlo}}]{Langford2017}%
  \BibitemOpen
  \bibfield  {author} {\bibinfo {author} {\bibfnamefont {N.~K.}\ \bibnamefont
  {Langford}}, \bibinfo {author} {\bibfnamefont {R.}~\bibnamefont
  {Sagastizabal}}, \bibinfo {author} {\bibfnamefont {M.}~\bibnamefont
  {Kounalakis}}, \bibinfo {author} {\bibfnamefont {C.}~\bibnamefont {Dickel}},
  \bibinfo {author} {\bibfnamefont {A.}~\bibnamefont {Bruno}}, \bibinfo
  {author} {\bibfnamefont {F.}~\bibnamefont {Luthi}}, \bibinfo {author}
  {\bibfnamefont {D.~J.}\ \bibnamefont {Thoen}}, \bibinfo {author}
  {\bibfnamefont {A.}~\bibnamefont {Endo}}, \ and\ \bibinfo {author}
  {\bibfnamefont {L.}~\bibnamefont {DiCarlo}},\ }\href {\doibase
  10.1038/s41467-017-01061-x} {\bibfield  {journal} {\bibinfo  {journal} {Nat.
  Commun.}\ }\textbf {\bibinfo {volume} {8}},\ \bibinfo {pages} {1715}
  (\bibinfo {year} {2017})}\BibitemShut {NoStop}%
\bibitem [{\citenamefont {Braum{\"u}ller}\ \emph {et~al.}(2017)\citenamefont
  {Braum{\"u}ller}, \citenamefont {Marthaler}, \citenamefont {Schneider},
  \citenamefont {Stehli}, \citenamefont {Rotzinger}, \citenamefont {Weides},\
  and\ \citenamefont {Ustinov}}]{Braumller2017}%
  \BibitemOpen
  \bibfield  {author} {\bibinfo {author} {\bibfnamefont {J.}~\bibnamefont
  {Braum{\"u}ller}}, \bibinfo {author} {\bibfnamefont {M.}~\bibnamefont
  {Marthaler}}, \bibinfo {author} {\bibfnamefont {A.}~\bibnamefont
  {Schneider}}, \bibinfo {author} {\bibfnamefont {A.}~\bibnamefont {Stehli}},
  \bibinfo {author} {\bibfnamefont {H.}~\bibnamefont {Rotzinger}}, \bibinfo
  {author} {\bibfnamefont {M.}~\bibnamefont {Weides}}, \ and\ \bibinfo {author}
  {\bibfnamefont {A.~V.}\ \bibnamefont {Ustinov}},\ }\href {\doibase
  10.1038/s41467-017-00894-w} {\bibfield  {journal} {\bibinfo  {journal} {Nat.
  Commun.}\ }\textbf {\bibinfo {volume} {8}},\ \bibinfo {pages} {779} (\bibinfo
  {year} {2017})}\BibitemShut {NoStop}%
\bibitem [{\citenamefont {Gherardini}\ \emph {et~al.}(2020)\citenamefont
  {Gherardini}, \citenamefont {Campaioli}, \citenamefont {Caruso},\ and\
  \citenamefont {Binder}}]{PhysRevResearch.2.013095}%
  \BibitemOpen
  \bibfield  {author} {\bibinfo {author} {\bibfnamefont {S.}~\bibnamefont
  {Gherardini}}, \bibinfo {author} {\bibfnamefont {F.}~\bibnamefont
  {Campaioli}}, \bibinfo {author} {\bibfnamefont {F.}~\bibnamefont {Caruso}}, \
  and\ \bibinfo {author} {\bibfnamefont {F.~C.}\ \bibnamefont {Binder}},\
  }\href {\doibase 10.1103/PhysRevResearch.2.013095} {\bibfield  {journal}
  {\bibinfo  {journal} {Phys. Rev. Research}\ }\textbf {\bibinfo {volume}
  {2}},\ \bibinfo {pages} {013095} (\bibinfo {year} {2020})}\BibitemShut
  {NoStop}%
\bibitem [{\citenamefont {Allahverdyan}\ \emph {et~al.}(2004)\citenamefont
  {Allahverdyan}, \citenamefont {Balian},\ and\ \citenamefont
  {Nieuwenhuizen}}]{Allahverdyan2004}%
  \BibitemOpen
  \bibfield  {author} {\bibinfo {author} {\bibfnamefont {A.~E.}\ \bibnamefont
  {Allahverdyan}}, \bibinfo {author} {\bibfnamefont {R.}~\bibnamefont
  {Balian}}, \ and\ \bibinfo {author} {\bibfnamefont {T.~M.}\ \bibnamefont
  {Nieuwenhuizen}},\ }\href {\doibase 10.1209/epl/i2004-10101-2} {\bibfield
  {journal} {\bibinfo  {journal} {Europhys. Lett.}\ }\textbf {\bibinfo {volume}
  {67}},\ \bibinfo {pages} {565} (\bibinfo {year} {2004})}\BibitemShut
  {NoStop}%
\bibitem [{\citenamefont {Ma}\ \emph {et~al.}(2021)\citenamefont {Ma},
  \citenamefont {Viennot}, \citenamefont {Kotler}, \citenamefont {Teufel},\
  and\ \citenamefont {Lehnert}}]{Ma2021}%
  \BibitemOpen
  \bibfield  {author} {\bibinfo {author} {\bibfnamefont {X.}~\bibnamefont
  {Ma}}, \bibinfo {author} {\bibfnamefont {J.~J.}\ \bibnamefont {Viennot}},
  \bibinfo {author} {\bibfnamefont {S.}~\bibnamefont {Kotler}}, \bibinfo
  {author} {\bibfnamefont {J.~D.}\ \bibnamefont {Teufel}}, \ and\ \bibinfo
  {author} {\bibfnamefont {K.~W.}\ \bibnamefont {Lehnert}},\ }\href {\doibase
  10.1038/s41567-020-01102-1} {\bibfield  {journal} {\bibinfo  {journal} {Nat.
  Phys.}\ }\textbf {\bibinfo {volume} {17}},\ \bibinfo {pages} {322} (\bibinfo
  {year} {2021})}\BibitemShut {NoStop}%
\bibitem [{\citenamefont {Juliusson}\ \emph {et~al.}(2016)\citenamefont
  {Juliusson}, \citenamefont {Bernon}, \citenamefont {Zhou}, \citenamefont
  {Schmitt}, \citenamefont {le~Sueur}, \citenamefont {Bertet}, \citenamefont
  {Vion}, \citenamefont {Mirrahimi}, \citenamefont {Rouchon},\ and\
  \citenamefont {Esteve}}]{PhysRevA.94.063861}%
  \BibitemOpen
  \bibfield  {author} {\bibinfo {author} {\bibfnamefont {K.}~\bibnamefont
  {Juliusson}}, \bibinfo {author} {\bibfnamefont {S.}~\bibnamefont {Bernon}},
  \bibinfo {author} {\bibfnamefont {X.}~\bibnamefont {Zhou}}, \bibinfo {author}
  {\bibfnamefont {V.}~\bibnamefont {Schmitt}}, \bibinfo {author} {\bibfnamefont
  {H.}~\bibnamefont {le~Sueur}}, \bibinfo {author} {\bibfnamefont
  {P.}~\bibnamefont {Bertet}}, \bibinfo {author} {\bibfnamefont
  {D.}~\bibnamefont {Vion}}, \bibinfo {author} {\bibfnamefont {M.}~\bibnamefont
  {Mirrahimi}}, \bibinfo {author} {\bibfnamefont {P.}~\bibnamefont {Rouchon}},
  \ and\ \bibinfo {author} {\bibfnamefont {D.}~\bibnamefont {Esteve}},\ }\href
  {\doibase 10.1103/PhysRevA.94.063861} {\bibfield  {journal} {\bibinfo
  {journal} {Phys. Rev. A}\ }\textbf {\bibinfo {volume} {94}},\ \bibinfo
  {pages} {063861} (\bibinfo {year} {2016})}\BibitemShut {NoStop}%
\bibitem [{\citenamefont {Bourassa}\ \emph {et~al.}(2012)\citenamefont
  {Bourassa}, \citenamefont {Beaudoin}, \citenamefont {Gambetta},\ and\
  \citenamefont {Blais}}]{PhysRevA.86.013814}%
  \BibitemOpen
  \bibfield  {author} {\bibinfo {author} {\bibfnamefont {J.}~\bibnamefont
  {Bourassa}}, \bibinfo {author} {\bibfnamefont {F.}~\bibnamefont {Beaudoin}},
  \bibinfo {author} {\bibfnamefont {J.~M.}\ \bibnamefont {Gambetta}}, \ and\
  \bibinfo {author} {\bibfnamefont {A.}~\bibnamefont {Blais}},\ }\href
  {\doibase 10.1103/PhysRevA.86.013814} {\bibfield  {journal} {\bibinfo
  {journal} {Phys. Rev. A}\ }\textbf {\bibinfo {volume} {86}},\ \bibinfo
  {pages} {013814} (\bibinfo {year} {2012})}\BibitemShut {NoStop}%
\bibitem [{\citenamefont {Schutjens}\ \emph {et~al.}(2013)\citenamefont
  {Schutjens}, \citenamefont {Dagga}, \citenamefont {Egger},\ and\
  \citenamefont {Wilhelm}}]{PhysRevA.88.052330}%
  \BibitemOpen
  \bibfield  {author} {\bibinfo {author} {\bibfnamefont {R.}~\bibnamefont
  {Schutjens}}, \bibinfo {author} {\bibfnamefont {F.~A.}\ \bibnamefont
  {Dagga}}, \bibinfo {author} {\bibfnamefont {D.~J.}\ \bibnamefont {Egger}}, \
  and\ \bibinfo {author} {\bibfnamefont {F.~K.}\ \bibnamefont {Wilhelm}},\
  }\href {\doibase 10.1103/PhysRevA.88.052330} {\bibfield  {journal} {\bibinfo
  {journal} {Phys. Rev. A}\ }\textbf {\bibinfo {volume} {88}},\ \bibinfo
  {pages} {052330} (\bibinfo {year} {2013})}\BibitemShut {NoStop}%
\bibitem [{\citenamefont {Peterer}\ \emph {et~al.}(2015)\citenamefont
  {Peterer}, \citenamefont {Bader}, \citenamefont {Jin}, \citenamefont {Yan},
  \citenamefont {Kamal}, \citenamefont {Gudmundsen}, \citenamefont {Leek},
  \citenamefont {Orlando}, \citenamefont {Oliver},\ and\ \citenamefont
  {Gustavsson}}]{PhysRevLett.114.010501}%
  \BibitemOpen
  \bibfield  {author} {\bibinfo {author} {\bibfnamefont {M.~J.}\ \bibnamefont
  {Peterer}}, \bibinfo {author} {\bibfnamefont {S.~J.}\ \bibnamefont {Bader}},
  \bibinfo {author} {\bibfnamefont {X.}~\bibnamefont {Jin}}, \bibinfo {author}
  {\bibfnamefont {F.}~\bibnamefont {Yan}}, \bibinfo {author} {\bibfnamefont
  {A.}~\bibnamefont {Kamal}}, \bibinfo {author} {\bibfnamefont {T.~J.}\
  \bibnamefont {Gudmundsen}}, \bibinfo {author} {\bibfnamefont {P.~J.}\
  \bibnamefont {Leek}}, \bibinfo {author} {\bibfnamefont {T.~P.}\ \bibnamefont
  {Orlando}}, \bibinfo {author} {\bibfnamefont {W.~D.}\ \bibnamefont {Oliver}},
  \ and\ \bibinfo {author} {\bibfnamefont {S.}~\bibnamefont {Gustavsson}},\
  }\href {\doibase 10.1103/PhysRevLett.114.010501} {\bibfield  {journal}
  {\bibinfo  {journal} {Phys. Rev. Lett.}\ }\textbf {\bibinfo {volume} {114}},\
  \bibinfo {pages} {010501} (\bibinfo {year} {2015})}\BibitemShut {NoStop}%
\bibitem [{\citenamefont {Pirkkalainen}\ \emph {et~al.}(2013)\citenamefont
  {Pirkkalainen}, \citenamefont {Cho}, \citenamefont {Li}, \citenamefont
  {Paraoanu}, \citenamefont {Hakonen},\ and\ \citenamefont
  {Sillanp{\"a}{\"a}}}]{Pirkkalainen2013}%
  \BibitemOpen
  \bibfield  {author} {\bibinfo {author} {\bibfnamefont {J.-M.}\ \bibnamefont
  {Pirkkalainen}}, \bibinfo {author} {\bibfnamefont {S.~U.}\ \bibnamefont
  {Cho}}, \bibinfo {author} {\bibfnamefont {J.}~\bibnamefont {Li}}, \bibinfo
  {author} {\bibfnamefont {G.~S.}\ \bibnamefont {Paraoanu}}, \bibinfo {author}
  {\bibfnamefont {P.~J.}\ \bibnamefont {Hakonen}}, \ and\ \bibinfo {author}
  {\bibfnamefont {M.~A.}\ \bibnamefont {Sillanp{\"a}{\"a}}},\ }\href {\doibase
  10.1038/nature11821} {\bibfield  {journal} {\bibinfo  {journal} {Nature
  (London)}\ }\textbf {\bibinfo {volume} {494}},\ \bibinfo {pages} {211}
  (\bibinfo {year} {2013})}\BibitemShut {NoStop}%
\bibitem [{\citenamefont {Tang}\ \emph {et~al.}(2020)\citenamefont {Tang},
  \citenamefont {Wu}, \citenamefont {Wang}, \citenamefont {Sun}, \citenamefont
  {Tang}, \citenamefont {Zhang}, \citenamefont {Li}, \citenamefont {Lu},
  \citenamefont {Xiao},\ and\ \citenamefont {Xia}}]{PhysRevA.101.053802}%
  \BibitemOpen
  \bibfield  {author} {\bibinfo {author} {\bibfnamefont {J.}~\bibnamefont
  {Tang}}, \bibinfo {author} {\bibfnamefont {Y.}~\bibnamefont {Wu}}, \bibinfo
  {author} {\bibfnamefont {Z.}~\bibnamefont {Wang}}, \bibinfo {author}
  {\bibfnamefont {H.}~\bibnamefont {Sun}}, \bibinfo {author} {\bibfnamefont
  {L.}~\bibnamefont {Tang}}, \bibinfo {author} {\bibfnamefont {H.}~\bibnamefont
  {Zhang}}, \bibinfo {author} {\bibfnamefont {T.}~\bibnamefont {Li}}, \bibinfo
  {author} {\bibfnamefont {Y.}~\bibnamefont {Lu}}, \bibinfo {author}
  {\bibfnamefont {M.}~\bibnamefont {Xiao}}, \ and\ \bibinfo {author}
  {\bibfnamefont {K.}~\bibnamefont {Xia}},\ }\href {\doibase
  10.1103/PhysRevA.101.053802} {\bibfield  {journal} {\bibinfo  {journal}
  {Phys. Rev. A}\ }\textbf {\bibinfo {volume} {101}},\ \bibinfo {pages}
  {053802} (\bibinfo {year} {2020})}\BibitemShut {NoStop}%
\bibitem [{\citenamefont {Chen}\ \emph {et~al.}(2018)\citenamefont {Chen},
  \citenamefont {Zhang},\ and\ \citenamefont {Xue}}]{PhysRevA.98.052314}%
  \BibitemOpen
  \bibfield  {author} {\bibinfo {author} {\bibfnamefont {T.}~\bibnamefont
  {Chen}}, \bibinfo {author} {\bibfnamefont {J.}~\bibnamefont {Zhang}}, \ and\
  \bibinfo {author} {\bibfnamefont {Z.-Y.}\ \bibnamefont {Xue}},\ }\href
  {\doibase 10.1103/PhysRevA.98.052314} {\bibfield  {journal} {\bibinfo
  {journal} {Phys. Rev. A}\ }\textbf {\bibinfo {volume} {98}},\ \bibinfo
  {pages} {052314} (\bibinfo {year} {2018})}\BibitemShut {NoStop}%
\bibitem [{\citenamefont {Zeytino\ifmmode~\breve{g}\else \u{g}\fi{}lu}\ \emph
  {et~al.}(2015)\citenamefont {Zeytino\ifmmode~\breve{g}\else \u{g}\fi{}lu},
  \citenamefont {Pechal}, \citenamefont {Berger}, \citenamefont {Abdumalikov},
  \citenamefont {Wallraff},\ and\ \citenamefont {Filipp}}]{PhysRevA.91.043846}%
  \BibitemOpen
  \bibfield  {author} {\bibinfo {author} {\bibfnamefont {S.}~\bibnamefont
  {Zeytino\ifmmode~\breve{g}\else \u{g}\fi{}lu}}, \bibinfo {author}
  {\bibfnamefont {M.}~\bibnamefont {Pechal}}, \bibinfo {author} {\bibfnamefont
  {S.}~\bibnamefont {Berger}}, \bibinfo {author} {\bibfnamefont {A.~A.}\
  \bibnamefont {Abdumalikov}}, \bibinfo {author} {\bibfnamefont
  {A.}~\bibnamefont {Wallraff}}, \ and\ \bibinfo {author} {\bibfnamefont
  {S.}~\bibnamefont {Filipp}},\ }\href {\doibase 10.1103/PhysRevA.91.043846}
  {\bibfield  {journal} {\bibinfo  {journal} {Phys. Rev. A}\ }\textbf {\bibinfo
  {volume} {91}},\ \bibinfo {pages} {043846} (\bibinfo {year}
  {2015})}\BibitemShut {NoStop}%
\end{thebibliography}%

\end{document}